\RequirePackage{fix-cm}
\documentclass[smalcondensed,natbib]{svjour3}                     
\smartqed  
\usepackage{graphicx}
\usepackage[T1]{fontenc}
\usepackage[utf8]{inputenc}

\usepackage{mathrsfs}
\usepackage[english]{babel}
\usepackage[font=small,labelfont=bf]{caption}
\usepackage[font=footnotesize,caption=false]{subfig}
\usepackage{etoolbox}
\usepackage{xcolor}[2007/01/21]
 \usepackage{multirow}
 \usepackage{amsmath}
\usepackage[ruled]{algorithm2e}

\SetAlFnt{\small}
\SetAlCapFnt{\small}
\SetAlCapNameFnt{\small}
\SetAlCapHSkip{0pt}
\IncMargin{-\parindent}

\usepackage{csquotes}
\usepackage{array, booktabs, caption}

\usepackage{makecell}

\DeclareMathSymbol{\mlq}{\mathord}{operators}{``}
\DeclareMathSymbol{\mrq}{\mathord}{operators}{`'}
\usepackage[garamond]{mathdesign}

\DeclareSymbolFont{cmcal}{OMS}{cmsy}{m}{n}
\SetSymbolFont{cmcal}{bold}{OMS}{cmsy}{b}{n}
\DeclareSymbolFontAlphabet{\mathcal}{cmcal}
\usepackage{mathptmx}      

 \usepackage{hyperref}
\usepackage[
 open,
 openlevel=3,
 atend
 ]{bookmark}[2011/12/02]

 \bookmarksetup{color=blue}
\begin{document}

\title{A Survey of Cellular Automata: Types, Dynamics, Non-uniformity and Applications (Draft version)\thanks{This research is partially supported by Innovation in Science Pursuit for
Inspired Research (INSPIRE) under Dept. of Science and Technology,
Govt. of India.}
}

\titlerunning{A Survey of Cellular Automata: Types, Dynamics, Non-uniformity and Applications}        

\author{Kamalika Bhattacharjee \and Nazma Naskar \and Souvik Roy \and Sukanta Das}


\institute{Kamalika Bhattacharjee, Souvik Roy {and} Sukanta Das,  
			\at
              Department of Information Technology, Indian Institute of Engineering and Science Technology, Shibpur, Howrah 711103, India;
              \email{kamalika.it@gmail.com, svkr89@gmail.com and sukanta@it.iiests.ac.in}           
           \and
          Nazma Naskar
          \at School of Computing Engineering, KIIT University, Bhubaneswar, Odisha, India\\
          \email{naskar.preeti@gmail.com }
}

\date{Received: date / Accepted: date}

\maketitle

\begin{abstract}
Cellular automata (CAs) are dynamical systems which exhibit complex global behavior from
simple local interaction and computation. Since the inception of cellular automaton (CA) by von Neumann in $1950s$, it has attracted the attention of several researchers over various backgrounds and fields for modelling different physical, natural as well as real-life phenomena. Classically, CAs are uniform. However, non-uniformity has also been introduced in update pattern, lattice structure, neighborhood dependency and local rule. In this survey, we tour to the various types of CAs introduced till date, the different characterization tools, the global behaviors of CAs, like universality, reversibility, dynamics etc. Special attention is given to non-uniformity in CAs and especially to non-uniform elementary CAs, which have been very useful in solving several real-life problems.
\end{abstract}

\section{Introduction}

From the end of the first half of $20^{th}$ century, a new approach has started to come in scientific studies, which after questioning the so called Cartesian analytical approach, says that interconnections among the elements of a system, be it physical, biological, artificial or any other, greatly effect the behavior of the system. In fact, according to this approach, knowing the parts of a system, one can not properly understand the system as a whole. In physics, David \cite{bohm2002wholeness} is one of the advocates of this approach. This approach is adopted in psychology by Jacob Moreno [see as example \cite{moreno1932application}], which later gave birth to a new branch of science, named Network Science. During this time, however, a number of models, respecting this approach, have started to be proposed [see, as an example, \cite{McCulloch1943}]. Cellular Automata (CAs) are one of the most important developments in this direction.

The journey of CAs was initiated by John von  \cite{Neuma66} for the modelling of biological \emph{self-reproduction}. A cellular automaton (CA) is a discrete dynamical system comprising of an orderly network of cells, where each cell is a finite state automaton. The next state of the cells are decided by the states of their neighboring cells following a local update rule. John von Neumann's CA is an infinite $2$-dimensional square grid, where each square box, called cell, can be in any of the possible $29$ states. The next state of each cell depends on the state of itself and its four neighbors. This CA can not only model biological self-reproduction, but is also computationally universal. The beauty of a CA is that simple local interaction and computation of cells results in a huge complex behavior when the cells act together.

Since their inception, CAs have captured the attention of a good number of researchers of diverse fields. They have been flourished in different directions -- as universal constructors, see for example \cite{Thatcher,banks1970universality}; models of physical systems, see as example \cite{Chopard}; parallel computing machine, see \cite{Toffo87}; etc. In recent years, the CAs are highly utilized as solutions to many real life problems. As a consequence, a number of variations of the CAs have been developed. In this survey, we will conduct a tour to a few aspects of CAs. Our purpose is not to explore any specific aspect of CAs in depth, rather to cover different directions of CAs research along with a good number of references. Interested readers may go to the appropriate references to explore their interests in detail.

We conduct this survey with respect to six categories. First, we tour to the various types of CAs, developed since their inception (Section~\ref{sscn_typeCA}). For example, von Neumann's CA and Wolfram's CA are of different types -- first one is defined over $2$-dimensional grid having $29$ states per cell, whereas latter is one-dimensional binary CA. Second, to study the behavior of CAs, few tools and parameters are developed. They are visited in Section~\ref{scn_chrtool}. However, the most interesting property of CAs is probably their complex global behavior due to very simple local computations of the cells. We inspect some of the most explored global behavior in Section~\ref{sscn_prpCA}.

During last two decades, a new direction in CAs research has been opening up, which introduces non-uniformity in CA structure. Practically, CAs are uniform in all respects - uniformity in cell's update (that is, synchronicity), uniformity in neighborhood dependency, and uniformity in local rule. Though the \emph{uniform} CAs are very good in modelling physical systems, researchers have introduced \emph{non-uniformity} in CA structure to model some physical systems in a better way, and most importantly, to solve some real life problems efficiently. We survey various aspects of non-uniformity in Section~\ref{scn_nunCA}, and put our deeper attention on a special type of non-uniform CAs in Section~\ref{scn_eca}.

Nowadays, CAs are not only modelling tools, but also technologies. Use of CAs in different application domains has established this fact in the last two decades. We explore some of such applications in Section~\ref{scn_appCA}.

\section{Types of Cellular Automata}
\label{sscn_typeCA}
In this section, we survey various types of CAs proposed till date. Before stating these types, let us define cellular automata formally.

\subsection{Basic definition}

A cellular automaton consists of a set of cells which are arranged as a regular network. Each cell of a CA is a finite automaton that uses a finite state set $\mathcal{S}$. The CAs evolve in discrete time and space. During evolution, a cell of a CA changes its state depending on the present states of its neighbors. That is, to update its state, a cell uses a {\em next state function}, also known as {\em local rule}, whose arguments are the present states of the cell's neighbors.
Collection of the states of all cells at a given time is called {\em configuration} of the CA. During evolution, a CA, therefore, hops from one configuration to another. 


\begin{definition}
\label{Def:basic}
A cellular automaton (CA) is a quadruple ($\mathscr{L},\mathcal{S},\mathcal{N},f$), where
\begin{itemize}
\item $\mathscr{L} \subseteq \mathbb{Z}^{D}$ is the $D$-dimensional cellular space. A set of cells are placed in the locations of $\mathscr{L}$ to form a regular network.

\item $\mathcal{S}$ is the finite state set.

\item $\mathcal{N} = (\overrightarrow{v_1}, \overrightarrow{v_2}, \cdots, \overrightarrow{v_N})$ is the neighborhood vector of $N$ distinct elements of $\mathscr{L}$ which associates one cell to its \emph{neighbors}. Generally, the neighbors of a cell are the nearest cells surrounding the cell. However, when the neighborhood vector $\mathcal{N}$ is given, then the neighbors of a cell at location $\overrightarrow{v} \in \mathscr{L}$ are at locations $(\overrightarrow{v} + \overrightarrow{v_i}) \in \mathscr{L}$, for all $i\in \{1, 2, \cdots, N\}$.

\item  $f:\mathcal{S}^{N}\rightarrow \mathcal{S}$ is called the local rule of the automaton. The next state of a cell is given by $f(a_1, a_2, \cdots, a_N)$, where $a_1, a_2, \cdots, a_N$ are the state of its $N$ neighbors.
\end{itemize}
\end{definition}

A cellular automaton is also identified by its {\em global transition function}. Let us consider $\mathcal{C}$ represents $\mathcal{S}^{\mathscr{L}}$, the set of all configurations. Then, a CA is a function $G:\mathcal{C}\rightarrow \mathcal{C}$, which is called global transition function. Classically, CAs are synchronous (that is, all the cells are updated simultaneously) and homogeneous (that is, all the cells use single next sates function). For these cases, if a configuration $y=(y_{\overrightarrow{v}})_{\overrightarrow{v} \in \mathscr{L}}$ is successor of another configuration $x=(x_{\overrightarrow{v}})_{\overrightarrow{v} \in \mathscr{L}}$, that is, $y=G(x)$, then $y$ is the result of the following application: \\
For each $\overrightarrow{v} \in \mathscr{L}$
\begin{equation}
y_{\overrightarrow{v}} = G(x)_{\overrightarrow{v}} = G(x_{\overrightarrow{v}})= f(x_{\overrightarrow{v}+\overrightarrow{v_1}}, x_{\overrightarrow{v}+\overrightarrow{v_2}}. \cdots, x_{\overrightarrow{v}+\overrightarrow{v_N}})
\end{equation}

Nevertheless, a CA is described in terms of four parameters -- $\mathscr{L}, \mathcal{S}, \mathcal{N}$ and $f$. Over the years, various types of CAs have been proposed varying the properties of these parameters. Even this basic definition of CA (Definition \ref{Def:basic}) is sometime compromised to get new types of CAs. In this section, we survey different types of CAs, developed till date, under the following headings --

\begin{enumerate}
\item The dimension and neighborhood of a cell.
\item The states of the cell.
\item The lattice size and boundary condition.
\item The local rule.
\end{enumerate}

\subsection{The dimension and neighborhood of a cell}
\label{Sec:DN}
\begin{figure}
     \subfloat[One way CA\label{1way}]{%
       \includegraphics[width=0.20\textwidth, scale = 0.7]{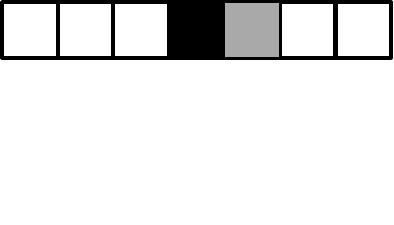}
     }
      \hfill
     \subfloat[Two way CA \label{2way}]{%
       \includegraphics[width=0.20\textwidth, scale = 0.7]{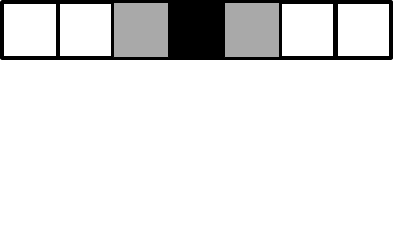}
     }
      \hfill
	\subfloat[von Neumann neighborhood \label{von}]{%
       \includegraphics[width=0.20\textwidth, scale = 0.7]{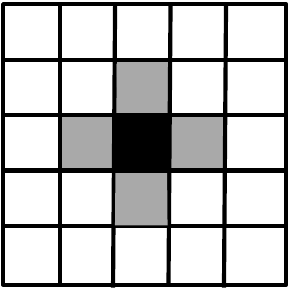}
     }
      \hfill
     \subfloat[Moore neighborhood \label{moore}]{%
            \includegraphics[width=0.20\textwidth, scale = 0.7]{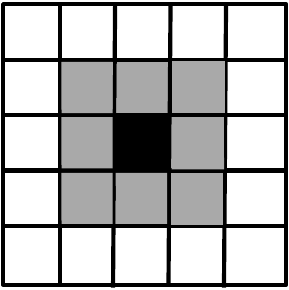}
     }
     \caption{Neighborhood dependencies: \ref{1way} and \ref{2way} are $1$-dimensional CAs, and \ref{von} and \ref{moore} are $2$-dimensional CAs. The black cell is the cell under consideration, and its neighbors are the shaded cells and the black cell itself}
     \label{fig:neighborhood}
 \end{figure}

The neighborhood of CA cells is strongly correlated with the dimension of the CAs. The original CA, proposed by von Neumann, is of $2$-dimension and uses $5$-neighborhood (orthogonal ones and itself) dependency. Fig.~\ref{von} shows such a dependency: the neighbors of the black cell is itself and the four shaded cells. This neighborhood dependency is traditionally known as {\em von Neumann neighborhood}.

A natural extension of this neighborhood dependency is the 9-neighborhood dependency, where four non-orthogonal cells are additionally considered as neighbors. Fig.~\ref{moore} shows this kind of neighborhood dependency, which was proposed by \cite{moore1962machine}, and is traditionally known as {\em Moore neighborhood}.
This neighborhood structure has been utilized to design the famous \emph{Game of Life}, a CA which was introduced by John Conway and popularized by Martin \cite{Gardner71}. 

\begin{figure}
   \subfloat[Margolus neighborhood\label{Fig:Margolus}]{%
       \includegraphics[scale = 0.3]{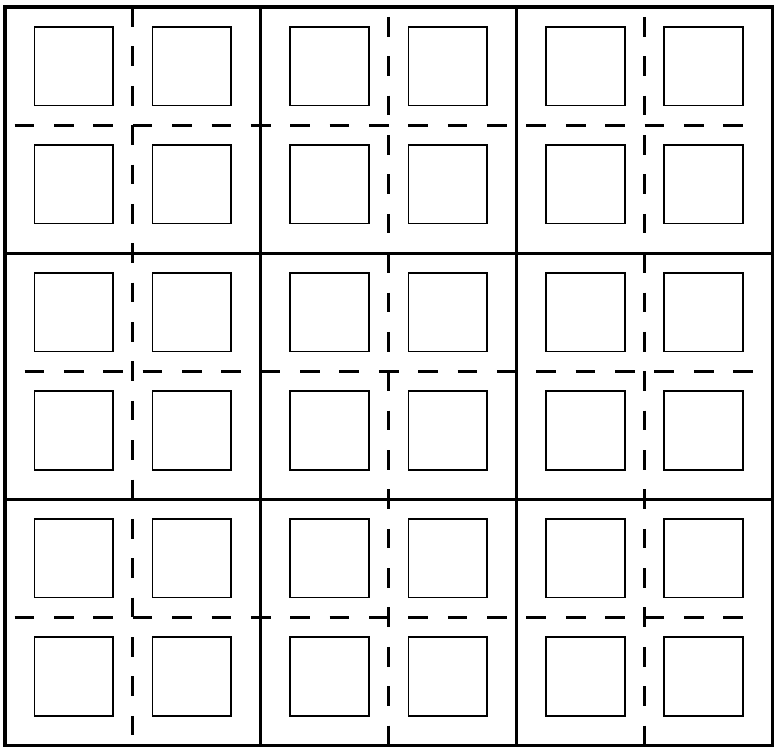}
     }
      \hfill
     \subfloat[Hexagonal CA\label{Fig:Hexagonal}]{%
       \includegraphics[scale = 0.3]{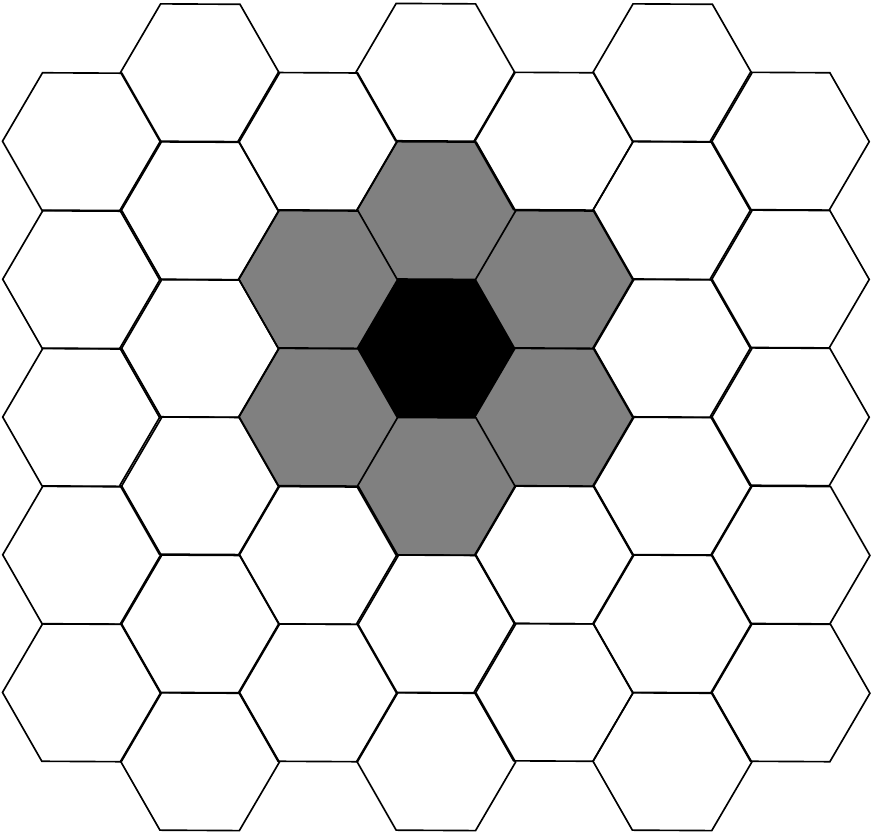}
     }
      \hfill
     \subfloat[Partitioned CA\label{Fig:PartitionedCA}]{%
       \includegraphics[scale = 0.3]{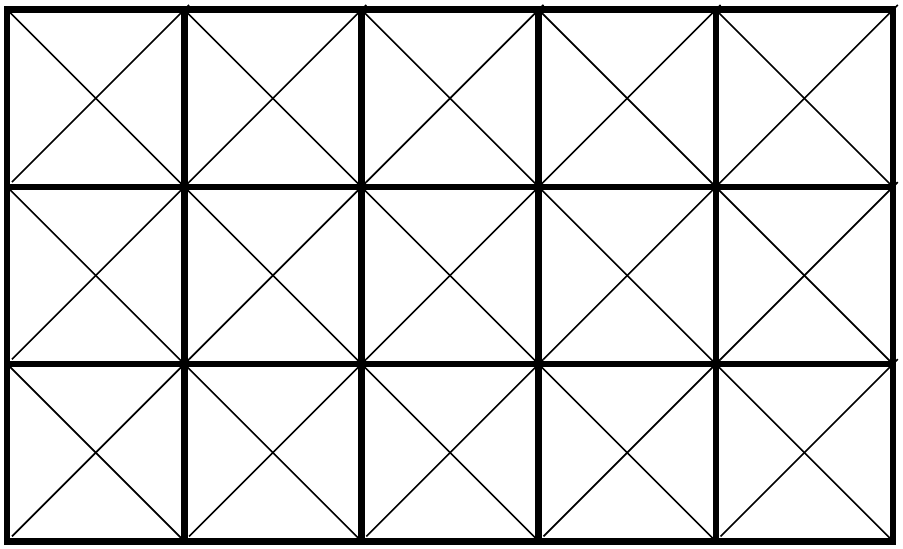}
     }

     \caption{Neighborhood dependencies: \ref{Fig:Margolus} is a block CA, \ref{Fig:Hexagonal} is hexagonal CA, and \ref{Fig:PartitionedCA} is a partitioned CA, where a square cell is partitioned into four triangular parts}
     \label{Fig:Margo+Hexa}
 \end{figure}

Apart from these two popular neighborhood dependencies for $2$-dimensional CAs, some other variations, such as Margolus neighborhood (\cite{Toffo87}) are also reported. In this neighborhood, the lattice $\mathscr{L}$ is divided into $2 \times 2$ non-overlapping blocks ($2 \times 2 \times 2$ cubes in three dimensions). The partitioning of $\mathscr{L}$ into blocks is applied on different spacial co-ordinates on alternate time steps. Fig.~\ref{Fig:Margolus} clarifies the idea: the smaller boxes are the cells, whereas the squares enclosing 4 cells are the blocks. There are two types of overlapping blocks, shown by dark lines and dotted lines. The cells of a block are neighbors of each other, and the type of blocks alternates in alternating time steps. This kind of CAs is also known as {\em block cellular automata} or {\em partitioning cellular automata}. In these CAs, the local rule, or rather {\em block rule}, updates the entire block as a distinct entity rather than any individual cell.

However, in the above types of CAs, the cells are considered as squares. In fact, in most of the CAs, the cells are square in shape. In some of the works, on the other hand, the cells are considered as hexagons over 2-dimensional space, and as a result, we get a different neighborhood dependency; see for example \cite{refId0,Siap2d}. Fig.~\ref{Fig:Hexagonal} shows such a neighborhood dependency. Not only hexagonal cells, \cite{IMAI2000181,morita2016universality} have worked with triangular cells also. Further, CAs have been defined in hyperbolic plane by \cite{margenstern1999polynomial,Margenstern200199}. As a result, we have been a witness of several types of CAs depending on the variations of neighborhood dependencies. As an extension of uniform neighborhood dependencies of cells of CAs, {\em Automata Networks} (see Section~\ref{sscn_NA}) have been defined over arbitrary networks, where the neighborhood dependencies of different cells may be different; see for example \cite{Marr20,tomassini29,Darabos7}.

{\em Partitioned cellular automata}, which are closely related to block cellular automata, are the CAs where each cell is partitioned into a finite number of parts. Hexagonal CA of \cite{refId0}, triangular CA of \cite{morita2016universality} are the partitioned CAs. Many properties of these CAs have been explored by Morita and his co-researchers. Fig.~\ref{Fig:PartitionedCA} shows the partitioning of square cells into four parts. Obviously, neighborhood dependencies of the parts are different than that of the cells. The local rules of partitioned CAs consider the states of parts during the state updates.

Sometimes, neighborhood of a CA is represented by \emph{radius}. By \emph{radius}, we mean the number of consecutive cells in a direction on which a cell depends on. For example, for $1$-dimensional CA, neighborhood $\mathcal{N}$ can be represented as $\mathcal{N}=\lbrace -r, -(r-1), \cdots, -1, 0, 1, \\\cdots, (r-1), r\rbrace$, where $r$ is the radius of the CA. In Fig.\ref{2way}, $r=1$. Classically, the radius is same in every direction.
However, in some works this restriction is removed; see for example \cite{Jump74,boccara-1998-31}. One of such CAs, called \emph{one-way CA}, has been proposed by \cite{Dyer80}, in which communication is allowed only in one direction, that is, in a $1$-dimensional array, the next state of each cell depends on itself and either of its left neighbor(s) or right neighbor(s) (see Fig.~\ref{1way}).

Although CAs are originally defined over 2-dimensional space, a large number of researches have been dedicated to explore one-dimensional CAs; \cite{Amoroso72,Pries86,suttner91,biplabtcad} are to name a few. \cite{Wolfr83,Wolframbook} has introduced a class of very simple CAs, called \emph{Elementary cellular automata} or ECAs, which are one-dimensional, two-state, and having three-neighborhood dependency (like Fig.~\ref{2way}). In last three decades, the major share of cellular automata research has gone to ECAs and their variations.

However, several fundamental properties and parameters of $2$-dimensional CAs are explored by \cite{Packa85b,Kari90,Durand93,terrier2004two,deOliveira20061,chr2D}, and many others. Higher than $2$-dimensional CAs are also proposed, see for example  \cite{gandin19973d,Palas1,miller2005two,Mo201431-3DCA}. For instance, in \cite{miller2005two} two-state three-dimensional reversible CA (RCA) is described and is shown to accomplish universal computation and construction. However, two or higher dimensional CAs sometime behave differently than $1$-dimensional CAs. For example, reversibility, an well addressed problem of $1$-D CA, is undecidable for $2$ or higher dimensional CAs, see \cite{Kari90}. In fact, most of the decision problems related to two or higher dimensional CAs are undecidable, see for example \cite{DENNUNZIO201440}.


\subsection{The states of the cell} 
A cell can be in any of the states of a finite state set $\mathcal{S}$ at any point of time. The number of states of von Neumann's CA is 29. Initially, many researchers targeted to simplify von Neumann's CA by reducing the number of states. For instance, state count of CAs was reduced to eight by \cite{Codd68}. \cite{Thatcher} has shown construction and computational universality as well as self-reproducing ability of  
von Neumann's cellular space, whereas \cite{Arbib66} depicted a simple self-reproducing CA capable of universality. Then, \cite{Banks71} has proved the universal computability of $2$-state CA and provided a description of self-reproducing CA using only $4$ states. However, all these constructions are for $2$-dimensional CAs, and use $5$-neighborhood dependency (see Fig.~\ref{von}). Conway's Game of Life, on the other hand, uses 9-neighborhood dependency (see Fig.~\ref{moore}) and two states per cell. \cite{Gardner71} has proved that this binary CA is also computationally universal. In fact, many CAs, including the ECAs, are binary; that is, the cells use only two states. \cite{Smith71} has shown that neighborhood size and state-set cardinality of a CA are interrelated. A CA with higher neighborhood size can always be emulated by another CA that has lesser neighborhood size but higher number of states per cell, and the contrariwise. 

Generally, the cellular space $\mathscr{L}$ is infinite. Hence, the configurations which are the collections of states of all cells of $\mathscr{L}$ at different time, are also infinite. As is in von Neumann's CA, sometime a restriction is imposed on CAs: a \emph{quiescent state} $q \in \mathcal{S}$ is considered so that $f(q,q,\cdots,q) = q$, where $f$ is the local rule (see Definition~\ref{Def:basic}). This means, a cell whose neighbors are in quiescent states, remains in quiescent state. As a result, we get a new class of configurations, called {\em finite configurations}.
\begin{definition}
\label{Def:FiniteConf}
Let $q \in \mathcal{S}$ be a quiescent state of a CA ($\mathscr{L},\mathcal{S},\mathcal{N},f$), that is, $f(q,q,\cdots,q) = q$, and the cellular space $\mathscr{L}$ be infinite.
A configuration of the CA is called a {\em finite configuration} if all but a finite number of cells are in $q$. 
\end{definition}
Thus, if the initial configuration of the CA is a finite configuration, at every time step the new configuration remains as a finite configuration. Let us consider $\mathcal{C}_F$ represents the set of all finite configurations. Obviously, $\mathcal{C}_F \subset \mathcal{C}$, where $\mathcal{C}=\mathcal{S}^{\mathscr{L}}$, the set of all configurations. Then, the global transition function of the CA over $\mathcal{C}_F$, ${G}_F: \mathcal{C}_F \rightarrow \mathcal{C}_F$ is a restriction of ${G}$. This restriction is at least needed to simulate a CA in computer.

In a number of works, such as \cite{Martin84a,ppc1}, states of a CA are considered as elements of a finite field. \cite{vlsi00d,biplabtcad} have further considered the states as elements of an extension field. \cite{ito1983linear,CATTANEO2004249} have considered the state set as $\mathbb{Z}_m$ (the integers modulo $m$) in their works. 

In almost all cases, however, every cell of a CA uses the same state set. In {\em polygeneous} CAs, on the contrary, a CA may use different state sets, see \cite{Sarkar00}.

\subsection{The boundary condition}
\begin{figure}
   \subfloat[Null boundary\label{Fig:NullB}]{%
       \includegraphics[scale = 0.35]{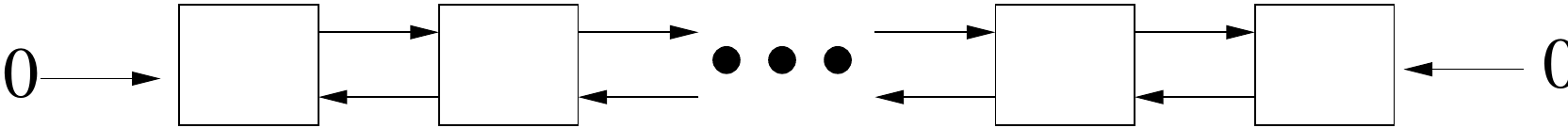}
     }
      \hfill
     \subfloat[Periodic boundary\label{Fig:PeriodicB}]{%
       \includegraphics[scale = 0.35]{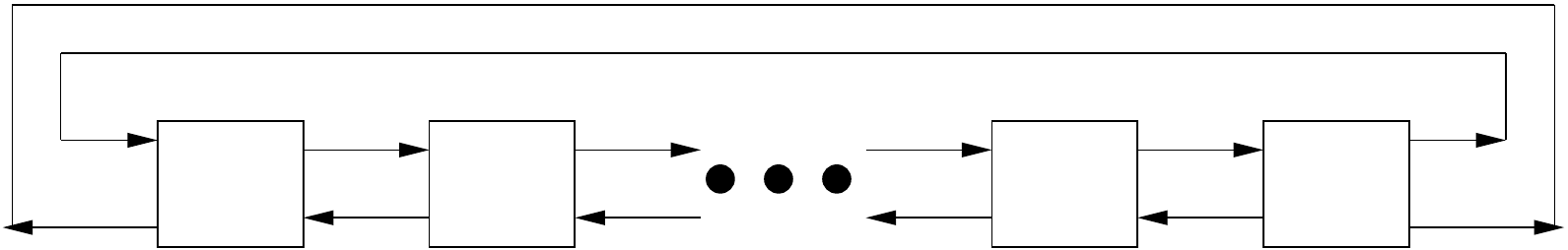}
     }
      \hfill
    \subfloat[Adiabatic boundary\label{Fig:AdiaB}]{%
       \includegraphics[scale = 0.35]{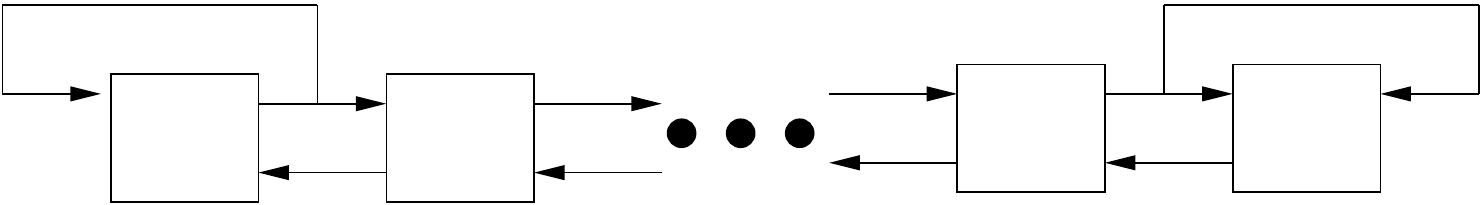}
     }
      \hfill
    \subfloat[Reflexive boundary\label{Fig:ReflB}]{%
       \includegraphics[scale = 0.35]{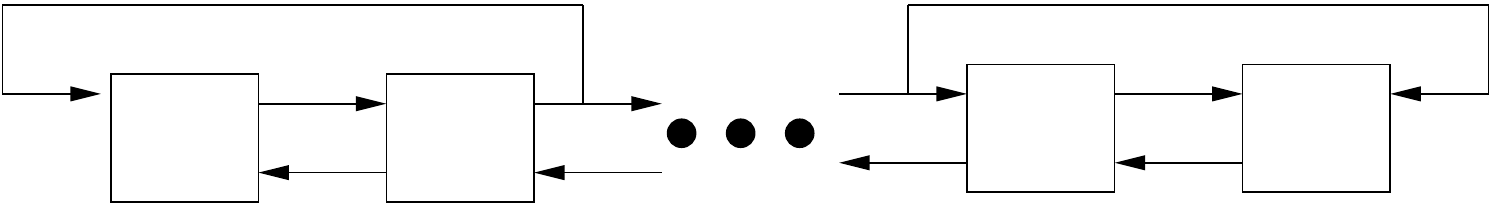}
     } 
           \hfill
    \subfloat[Intermediate boundary\label{Fig:IntmB}]{%
       \includegraphics[scale = 0.35]{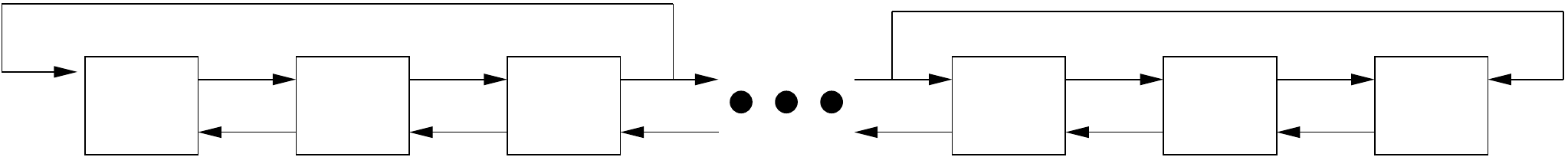}
     }     
     \caption{Boundary conditions of 1-D finite CAs. Arrows pointing to a cell indicate the dependencies of the cell. Here all the CAs use 3-neighborhood dependency}
     \label{Fig:BoundaryC}
 \end{figure}

As the cellular space $\mathscr{L}$ is usually infinite, there is no question of boundary, and boundary condition. However, in a few works, $\mathscr{L}$ is assumed as finite, which is obviously having boundaries. These CAs are finite CAs.
\begin{definition}
A CA is called as a {\em Finite Cellular Automaton} if the cellular space $\mathscr{L} \subseteq \mathbb{Z}^{D}$ is finite. Otherwise, it is an {\em Infinite Cellular Automaton}.
\end{definition}
Finite CAs are really important if the automata are to be implemented. Two boundary conditions for finite CAs are generally used -- open boundary condition and periodic boundary condition. 
In open (fixed) boundary CAs, the missing neighbors of extreme cells (leftmost and rightmost cells in case of $1$-D CAs) are usually assigned some fixed states. Among the open boundary conditions, the most popular is null boundary, where the missing neighbors of the terminal cells are always in state $0$ (Null). Null boundary CAs have received a good attention by the researchers, who have used CAs for VLSI (Very Large Scale Integration) design and test. \cite{Horte89a,Tsali91,ppc1,biplabtcad,tcad/DasS10} are some examples of such works. Fig.~\ref{Fig:NullB} explains null boundary condition of a 1-dimensional CA.

In case of periodic boundary condition, on the other hand, the boundary cells are neighbors of some other boundary cells. For instance, for $1$-D CAs, the rightmost and leftmost cells are neighbors of each other (see Fig.~\ref{Fig:PeriodicB}). Higher dimensional CAs under periodic boundary conditions have also been explored by some researchers, such as \cite{Palas1,Jin2012538,uguz2013reversibility} etc. However, the periodic boundaries of a 2-D CA can be visualized by folding right and left sides of a rectangle to form a tube, and then taping top and bottom edges of the tube to form a torus.

Periodic boundary conditions sometime relate finite CAs and infinite CAs through {\em periodic configurations}. 
\begin{definition}
A configuration $x\in \mathcal{C}$ of a D-dimensional infinite CA is {\em periodic} if the configuration is invariant under $D$ linearly independent translations. That is, there exist $D$ natural numbers $n_1, n_2,\cdots, n_D$ so that $x$ is same as $n_i$ times shift of $x$ in dimension $i$, for each $i\in\{1, 2, \cdots, D\}$.
\end{definition}
Let $\mathcal{C}_P\subset {\mathcal{C}}$ denote the set of all periodic configurations of a CA. Then, $G_P:\mathcal{C}_P \rightarrow \mathcal{C}_P$ is a restriction of $G$. However, the members of $\mathcal{C}_P$ preserve periodicity of configurations, which is also preserved in periodic boundary condition. So, periodic configurations are sometime called as the periodic boundary conditions of finite CAs.

Some other variations of boundary conditions also exist, such as adiabatic (or reflecting) boundary, reflexive boundary and intermediate boundary. In adiabatic (or reflecting) boundary condition, the boundary cells assume the missing neighbors as their duplicates. Fig.~\ref{Fig:AdiaB} shows this boundary condition where the left (right) neighbor of the leftmost (rightmost) cell is the cell itself.
In reflexive boundary condition, shown in Fig.~\ref{Fig:ReflB}, the left and right neighbors of boundary cells are the same cell. Finally, Fig.~\ref{Fig:IntmB} explains the intermediate boundary condition for a 1-D CA, where a missing neighbor of a boundary cell is the neighbor's neighbor of the boundary cell. \cite{Nandi96} have used the intermediate boundary condition to improve quality of pseudo-random patterns, generated by CAs.

As another special case of open boundary condition, stochastic boundary condition is proposed, where the boundary cells assume some states stochastically. \cite{PhysRevE-CKS} have used this boundary condition to model traffic flow on a one-dimensional lattice with finite number of cells. In this traffic model, the authors have considered that the cars are injected at the leftmost cell with a probability, and the cars are removed from the right boundary with another probability.

\subsection{Local rule} 
\label{Sec:rule}
A cell of a CA changes its state following a next state function $f:\mathcal{S}^{N}\rightarrow \mathcal{S}$ where $\mathcal{S}$ is the set of states and $N$ is the size of the neighborhood. The map $f$, which is generally known as the local rule of the CA, can be conveyed in different ways. As an instance, in Conway's Game of Life, the local rule is $f:\{0,1\}^{9}\rightarrow \{0,1\}$, where state $0$ represents a dead cell and state $1$ represents an alive cell. It is stated as following in \cite{mitcourseware}:

\begin{displayquote}
\begin{itemize}
\item ``Birth: a cell that is dead at time $t$, will be alive at time $t + 1$, if exactly $3$ of its eight neighbors were alive at time $t$.
\item Death: a cell can die by:
\begin{itemize}
\item Overcrowding: if a cell is alive at time $t$ and $4$ or more of its neighbors are also alive at time $t$, the cell will be dead at time $t + 1$.

\item Exposure: If a live cell at time $t$ has only $1$ live neighbor or no live neighbors, it will be dead at time $t + 1$.
\end{itemize}
\item Survival: a cell survives from time $t$ to time $t + 1$ if and only if $2$ or $3$ of its neighbors are alive at time $t$.''
 \end{itemize}
 \end{displayquote}

However, the local rule $f$ can also be represented in a tabular form. In this form, the table contains entries for the next state values corresponding to each of the possible neighborhood combinations according to the local rule. This representation has been popularized by \cite{Wolfr83}, and used to name his Elementary CAs (ECAs). Table~\ref{rt2} shows seven rules of ECAs, which are named after the decimal equivalents (last column of Table~\ref{rt2}) of 8-bit binary sequence. Obviously, the tabular form is good if size of neighborhood ($N$) and the state set ($\mathcal{S}$) are very small.

\begin{table}[h]
\setlength{\tabcolsep}{1.3pt}
\begin{center}
\caption{ECAs rules. Here, PS and NS represent present state and next state respectively}
\label{rt2}
\resizebox{0.50\textwidth}{!}{
\begin{tabular}{cccccccccc}
 \toprule
\thead{PS} \hspace{1em}& \thead{111} & \thead{110} & \thead{101} & \thead{100} & \thead{011} & \thead{010} & \thead{001} & \thead{000} & \hspace{1em} \multirow{3}{*}{\thead{Rule}}\\ 

\thead{(RMT)} \hspace{1em}& \thead{(7)} & \thead{(6)} & \thead{(5)} & \thead{(4)} & \thead{(3)} & \thead{(2)} & \thead{(1)} & \thead{(0)} & \\ 
 \midrule
\multirow{3}{*}{}
 &0 & 1 & 0 & 1 & 1 & 0 & 1 & 0 &\hspace{1em} 90\\
 & 1 & 0 & 0 & 1 & 0 & 1 & 1 & 0 &\hspace{1em} 150\\
 & 0& 0& 0& 1& 1& 1& 1& 0&\hspace{1em} 30\\
\thead{NS} & 0& 0& 0& 0& 0& 1& 0& 1&\hspace{1em} 5\\
 & 0& 1& 0& 0& 1& 0& 0& 1&\hspace{1em} 73\\
 & 1& 1& 0& 0& 1& 0& 0& 0&\hspace{1em} 200\\
 & 0& 1& 0& 1& 0& 0& 0& 0&\hspace{1em} 80\\
    \bottomrule
\end{tabular}
}
\end{center}
\end{table} 

Against a binary map $f:\{0,1\}^{N}\rightarrow \{0, 1\}$, an $N$-tuple $(s_1,s_2,\cdots,s_N) \in \mathcal{S}^{N}$ can be viewed as a {\em Min Term} of $N$-variable switching function. So, each of the tuples of the map $f$ is named as {\em Rule Min Term} (RMT). Table~\ref{rt2} shows eight RMTs against a rule. In the works of present authors, the RMTs of rules have been utilized. This concept is further extended by \cite{jca2015} to use it by non-binary local rules. However, all the above works that use RMTs are about 1-dimensional CAs.
\begin{definition}
\label{Def:RMT}
Let $f:\mathcal{S}^{N}\rightarrow \mathcal{S}$ be a local rule of a CA. A tuple $(s_1,s_2,\cdots,s_N) \in \mathcal{S}^{N}$ is called as a {\em Rule Min Term} (RMT) of the rule $f$, and is usually represented by it decimal equivalent. If $\mathcal{S}=\{0, 1,\cdots, d-1\}$, then the RMT is $r=s_1.d^{N-1}+s_2.d^{N-2}+\cdots+s_{N-1}.d+s_N$. The next state against this RMT of the rule, that is $f(s_1,s_2,\cdots,s_N)$, is also represented as $f[r]$.
\end{definition}

Therefore, a rule can be seen as a collection of RMTs along with their next states. In that sense, RMTs are atomic elements of a rule. The RMTs along with their next state values are sometime called as {\em transitions}. In case of ECAs, where $f:\{0,1\}^{3}\rightarrow \{0, 1\}$, a transition is a quadruplet $(x,y,z,f(x,y,z))$. A transition of an ECA rule is {\em active} if $f(x,y,z)\neq y$; otherwise the transition is {\em passive}. A rule $f$ can also be represented by its active transitions only. To present a rule, in general, at least its active transitions are to be noted down. The rule of Game of Life (see above), for example, shows all active transitions. In case of ECAs, although the numbering style of \cite{Wolfr83} (as shown in Table~\ref{rt2}) is widely used, they are alternatively identified by their active transitions only, known as {\em transition codes}. In the works of \cite{Fates20061,fates00608485}, for example, transition codes of ECAs rules have been used.

When the rule $f$ of a CA is linear, then the CA is also linear. That is, the global transition function $G$ of the CA is a linear function. In case of linear rule $f:\mathcal{S}^N\rightarrow \mathcal{S}$, the rule can be expressed as $f (a_1, a_2, \cdots, a_N) = \sum\limits_{i=1}^N c_i.a_i$, where $c_i$ $\in \mathcal{S}$ is a constant. And, the set $\mathcal{S}$ forms a commutative ring with identity.
There are seven linear ECAs for rules 60, 90, 102, 150, 170, 204 and 240 (excluding rule 0, which is also trivially linear). Apart from these ECAs, there is no additional additive ECAs. So, these rules are presented by some authors as {\em linear/additive} rules.

\begin{figure}
   \subfloat[Triangular Partitioned CA\label{Fig:TPCAr}]{%
       \includegraphics[scale = 0.4]{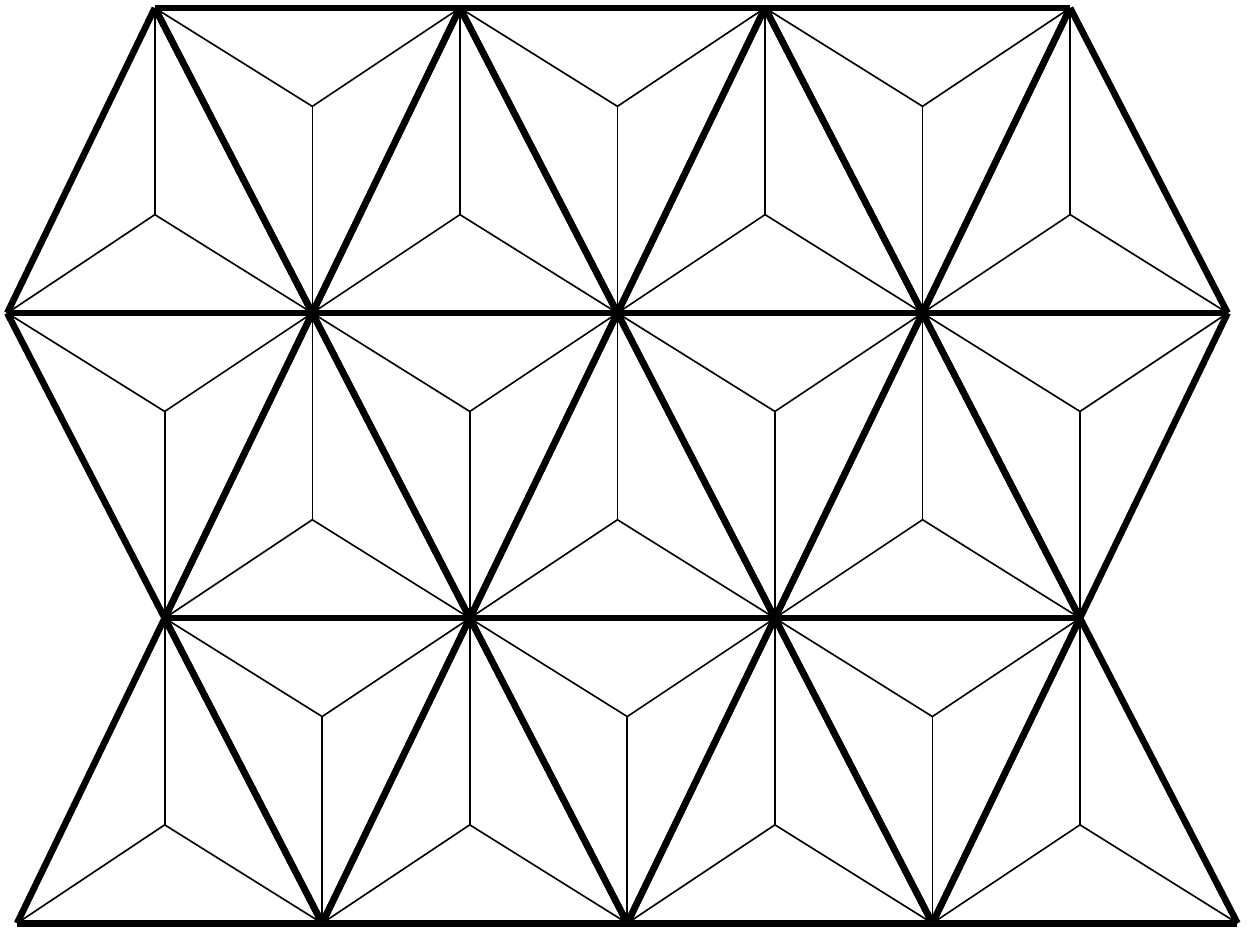}
     }
      \hfill
     \subfloat[A local rule\label{Fig:TPCArule}]{%
       \includegraphics[scale = 0.4]{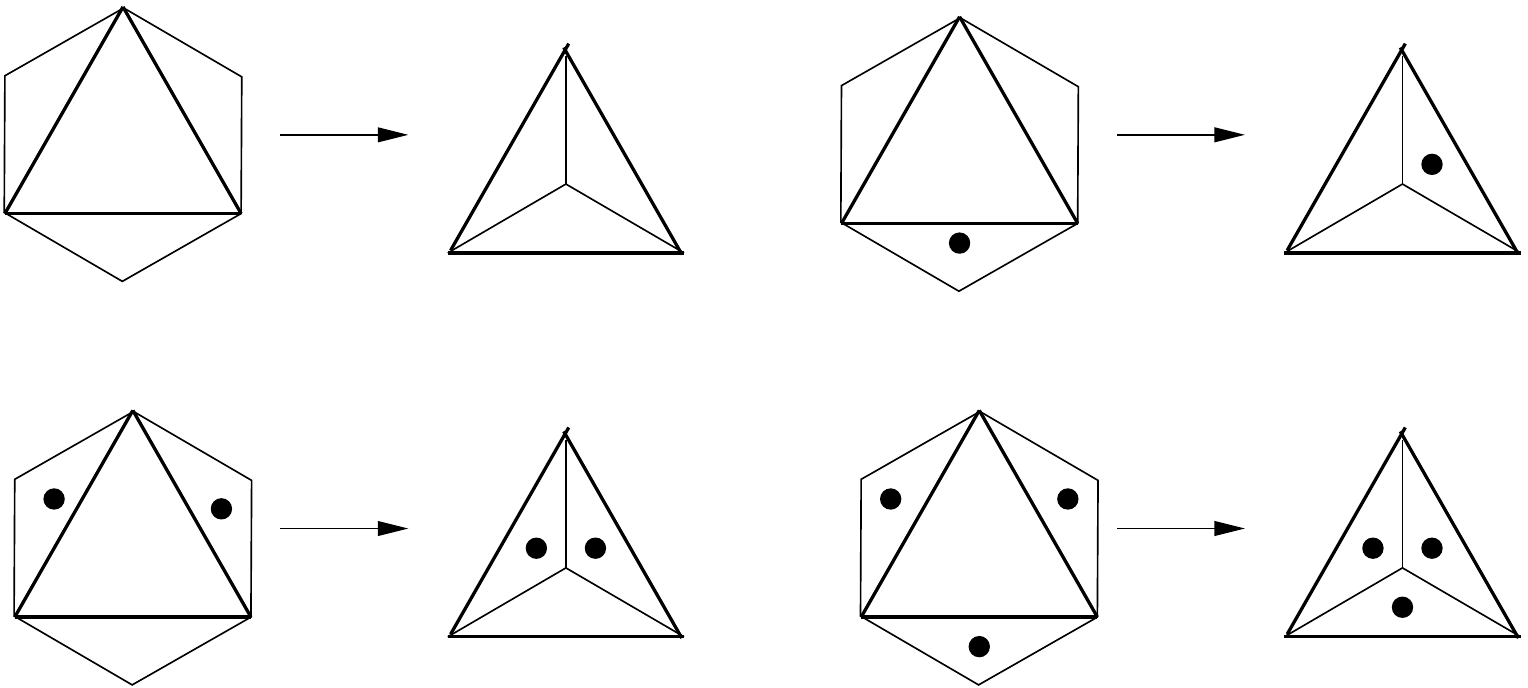}
     }    

     \caption{Triangular partitioned CA and its rule (as given by \cite{morita2016universality}). The rule is rotation symmetric}
     \label{Fig:TPCA+TPCArule}
 \end{figure}

In case of block cellular automata and partitioned cellular automata, which have been briefly discussed in Section~\ref{Sec:DN}, presentation of local rules is altogether different. The rule ({\em block rule}) of block CA does not change individual cells of the CA, rather it looks at the content of the block and updates the whole block. \cite{Toffo87} have described such rule in their book. For partitioned CAs, on the other hand, each cell is divided into several parts and the next state of each cell is determined by the states of the adjacent parts of the neighbor cells. Fig.~\ref{Fig:TPCA+TPCArule} shows a triangular partitioned CA and its rule, which is explored by \cite{morita2016universality}. Like block CA, the local rule of a partitioned CA updates all the parts of a cell. Unlike traditional CA, however, the local rule does not consider the present states of the neighbors of a cell, but adjacent parts of neighboring cells (Fig.~\ref{Fig:TPCArule}).

Classically, the CAs are uniform, synchronous and deterministic -- that is, all cells are updated together with the same local rule. A new class of non-classical CAs have been proposed where cells can follow different rules. These CAs are known as {\em hybrid} or {\em non-uniform} CAs. Historically, this class of CAs assume that the rules are the ECAs rules only. \cite{Pries86,Horte89a,ppc1}, and many others have explored these CAs, which are one-dimensional and finite. In this case, one needs to define the local rules for individual cells; hence she needs a rule vector $\mathcal{R}$.
\begin{definition}
\label{Def:RuleVector}
Let $\mathscr{L}= \mathbb{Z}/n\mathbb{Z}$ be the cellular space, where $n$, a natural number, represents the number of cells in the space.
A {\em rule vector} of a non-uniform CA over $\mathscr{L}$ is $\mathcal{R}=\langle \mathcal{R}_0, \mathcal{R}_1, \cdots, \mathcal{R}_{n-1}\rangle$, where $\mathcal{R}_i$ is a rule used by cell $i$, $i\in \mathscr{L}$.
\end{definition}
As a proof of concept, let us consider a 5-cell non-uniform CA with rule vector $\mathcal{R}=\langle 90, 150, 90, 5, 150\rangle$. This implies that the first and third cells use ECA rule 90, second and fifth cells use rule 150, and the fourth cell uses rule 5. This class of CAs are specially utilized in VLSI design and test.
Another type of CA, called a programmable CA (with respect to VLSI design), exists, where a cell can chose a distinct local rule at every time instant, see for example \cite{Nandi94a}. Nevertheless, a special kind of CAs have also been introduced where a CA updates itself in asynchronous way. In short, non-classical CAs have gained popularity in recent times. Section~\ref{scn_nunCA} and Section~\ref{scn_eca} are dedicated to discuss these non-classical CAs.

\section{Characterization tools of Cellular Automata}
\label{scn_chrtool}

In \emph{A New Kind of Science}, \cite{Wolframbook1} has argued that, to find out how a particular CA will behave, one has to observe what is happening just by running the CA. Predicting behavior of a system by means of (mathematical) analysis and without running it is only possible, according to Wolfram, for special systems with simple behavior (page $6$ of \cite{Wolframbook1}). In spite of this observation of some CAs researchers, a few characterization tools and parameters have been proposed in different time to analyze and predict the behavior of some CAs. Needless to say, all kind of dynamic behaviors of a CA may not be analyzed by a tool, but tools are used to discover some specific properties of a CA.

In this section, we survey the characterization tools and parameters, developed till date to analyze the CAs. Tools are mainly developed for one-dimensional CAs, and for two or higher dimensional CAs, ``run and watch'' is the primary technique to study the behavior. Though, few parameters are proposed which can be used to guess the behavior of two or higher dimensional CAs.

\subsection{De Bruijn graph}\label{dbg}
In the different development phases of CAs, graph theory has played an important role. One of the roles is describing the evolution of an automaton, and another is relating local properties to global properties. As an automaton has states which are mapped to another states using overlapping neighborhood sequence, it is very obvious to treat it by shift register. So, de Bruijn graph is considered as an alternative characterization tool.

\begin{definition}
Let $\Sigma$ be a set of symbols, and $s\ge 1$ be a number. Then, the {\em de Bruijn graph} is ${B(s, \Sigma)} = (V, E)$, where $V=\Sigma^s$ is the set of vertices, and $E=\{(ax,xb)|a, b\in \Sigma, x\in \Sigma^{s-1} \}$ is the set of edges.
\end{definition}

\begin{figure}
   \subfloat[The de Bruijn graph $B(2,\{0,1\})$\label{Fig:DeBruijn}]{%
       \includegraphics[scale = 0.5]{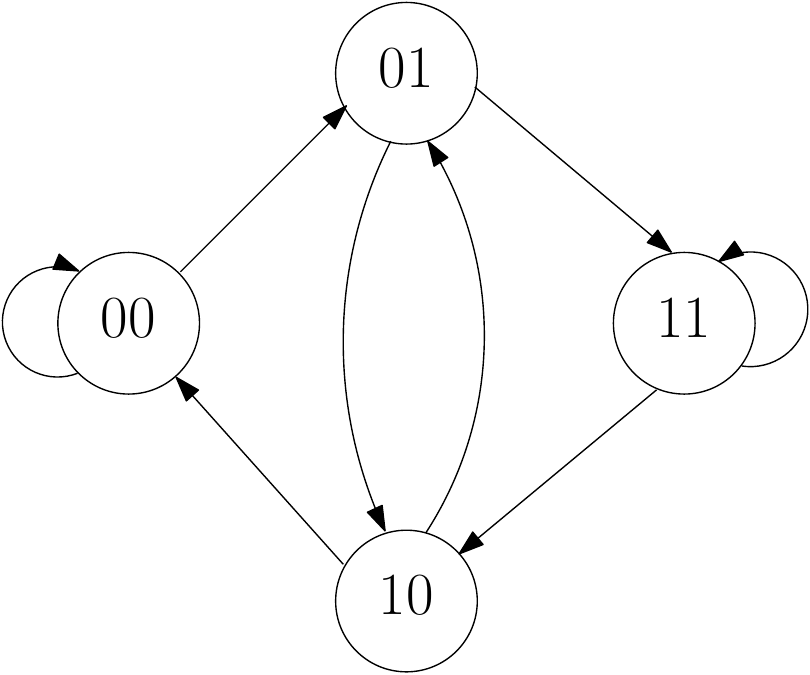}
     }
      \hfill
     \subfloat[De Bruijn graph for rule 90\label{Fig:DB90}]{%
       \includegraphics[scale = 0.5]{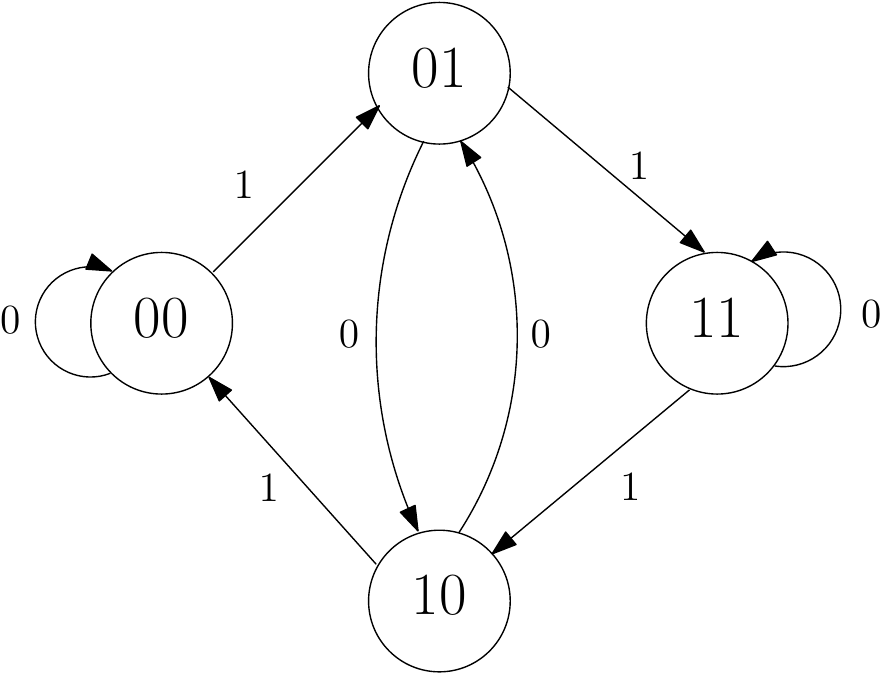}
     }    

     \caption{The de Bruijn Graph of CA with rule $90$ ($2^{nd}$ row of Table~\ref{rt2})}
     \label{Fig:DeBruijn90}
 \end{figure}

Fig.~\ref{Fig:DeBruijn} shows the de Bruijn graph $B(2,\{0,1\})$. A de Bruijn graph has $m^s$ vertices, consisting of all possible length-$ m $ sequences of $\Sigma$, where $m$ is the number of symbols in $\Sigma$. This graph is balanced in the sense that each vertex has both in-degree and out-degree $s$. 

A one-dimensional CA can be represented by a de Bruijn graph. Let us consider $s=N-1$ and $\Sigma=\mathcal{S}$, where $N$ is the size of neighborhood and $\mathcal{S}$ is the state-set of a 1-D CA (see Definition~\ref{Def:basic}). The edges $(ax,xb)$ of $B(N-1,\mathcal{S})$ show the overlap of nodes and $(axb)\in \mathcal{S}^N$. Now label each edge $(ax,xb)$ of $B(N-1,\mathcal{S})$ by $f(axb)\in \mathcal{S}$, where $f$ is the local rule. This labelled graph represents a one-dimensional CA with rule $f$, state set $\mathcal{S}$, and neighborhood size $N$. Since this graph does not relate to the lattice size, de Bruijn graph can be used to study finite and infinite 1-D CAs.

Fig.~\ref{Fig:DB90} is the de Bruijn graph for ECA rule 90 (whereas Fig.~\ref{Fig:DeBruijn} is a graph for any ECA).  This graph shows that if the left, self and right neighbors of a cell are all $0$s, then next state of the cell (that is, $f(0,0,0)$) is $0$, if the neighbors are $0, 0$ and $1$ respectively, the next state is $1$, and so on. In his work, \cite{suttner91} has defined {\em $s$-fusion operation}, which is equivalent to labelling an edge in above manner.

Over the years, this graph has been used by various researchers to understand global behavior, like surjectivity and reversibility, number conservation, equicontinuity etc. of 1-dimensional CAs, see for example \cite{suttner91,Soto2008,voorhees2008remarks}. In \cite{Martinez2008}, ECA $110$ has been explored to determine a glider-based regular expressions.  \cite{Mora2008} have studied cyclic properties and inverse of a CA using pair diagram of de Bruijn graph. 
\cite{Betel2013} have used de Bruijn graph to solve the parity problem. Recently, \cite{Mariot2017} have studied the periods of pre-images of spatially periodic configurations in surjective CA using de Bruijn graphs.

The de Bruijn graphs are traditionally used to represent and study classical CAs. However, in recent past, \cite{DennunzioFP13} have shown that de Bruijn graph can also be used to represent non-uniform CAs. 

\subsection{Matrix algebra}
\label{matrix}

Unlike de Bruijn graph, matrix algebra can represent only finite CAs. In fact, this characterization tool has been developed by \cite{Das90c} to study the behavior of 1-D hybrid or non-uniform CA that uses only linear ECA rules (see Section~\ref{Sec:rule}). A non-uniform CA is {\em linear/additive} if all of its local rules are linear/additive. So non-uniform finite linear/additive CAs can be characterized by matrix algebra. \cite{ppc1} have provided a good account of works on linear/additive non-uniform finite CAs in their book.

\begin{table}[h]
\begin{center}
\caption{Linear and complemented rules}
\label{CArule}
\resizebox{0.90\textwidth}{!}{
\begin{tabular}{ll|ll}
 \toprule
\thead{Rule} & \thead{Linear} & \thead{Rule} & \thead{Complement} \\
 & \thead{(With $XOR$ logic)} &  & \thead{(With $XNOR$ logic)} \\
 \midrule
					
 60: & $S_i(t+1)~=~S_{i-1}(t) \oplus S_{i}(t) $& 195: & $S_i(t+1)~=\overline{ ~S_{i-1}(t) \oplus S_{i}(t)} $\\
 90: & $S_i(t+1)~=~S_{i-1}(t) \oplus S_{i+1}(t)$& 165: & $S_i(t+1)~=\overline{~S_{i-1}(t) \oplus S_{i+1}(t)}$\\
 102: & $S_i(t+1)~=~S_{i}(t) \oplus S_{i+1}(t)$& 153: & $S_i(t+1)~=\overline{~S_{i}(t) \oplus S_{i+1}(t)}$\\
 150: & $S_i(t+1)~=~S_{i-1}(t) \oplus S_{i}(t) \oplus S_{i+1}(t)$ & 105: & $S_i(t+1)~=\overline{~S_{i-1}(t) \oplus S_{i}(t) \oplus S_{i+1}(t)}$\\
 170: & $S_i(t+1)~=~S_{i+1}(t)$ & 85: & $S_i(t+1)~=\overline{~S_{i+1}(t)}$\\
 204: & $S_i(t+1)~=~S_{i}(t) $ & 51: & $S_i(t+1)~=\overline{~S_{i}(t)} $\\
 240: & $S_i(t+1)~=~S_{i-1}(t) $ & 15: & $S_i(t+1)~=\overline{~S_{i-1}(t)} $\\
 
 \bottomrule
\end{tabular}
}
\end{center}
\end{table}

A binary linear rule $f$ can be expressed as XOR of input variables. For example, ECA rule 90 is a linear rule, and $f_{90}(x,y,z)=x\oplus z$. All the linear ECA rules can similarly be expressed by XOR logic. Table~\ref{CArule} shows such expressions, which is reproduced directly from \cite{ppc1}. However, if we look at the XORed expression of ECA rule 90, it is understood that a cell of ECA 90 depends on left and right neighbors only. When characterized by matrix algebra, a binary finite (uniform/non-uniform) CA is represented by a matrix, and then this neighborhood dependency is taken care of.

Let us consider a binary finite (uniform/non-uniform) CA having $n$ cells. This CA is represented by an $n \times n$ characteristics matrix operating on $GF(2)$ In this matrix, the $i^{th}$ row represents the dependency of the $i^{th}$ cell to its neighbors. The characteristics matrix ($T$), in this case, is formed as:

\begin{equation}
``
  \begin{array}{l}
   \mbox{$T\left[ i, j\right] $ =}
    \left\{
	\begin{array}{ll}
	    \mbox{1  if the next state of the $i^{th}$ cell depends on the present state of the $j^{th}$ cell}   ~~~~ \text{''}\\
	    \mbox{0  otherwise}
	    \end{array}\right.
    \end{array}
 \end{equation}  

The non-uniform CAs can use different rules in different cells, hence we need a {\em rule vector} to specify the rules against cells (see Definition~\ref{Def:RuleVector}). Let us take a $4$-cell non-uniform CA with rule vector $\mathcal{R}=\langle 150, 150, 90, 150 \rangle$  under null boundary condition. Then, the characteristics matrix of the CA is: 
\[
   T
=
\begin{bmatrix}
    1 & 1 & 0 & 0 \\
    1 & 1 & 1 & 0 \\
    0 & 1 & 0 & 1 \\
    0 & 0 & 1 & 1
\end{bmatrix}
\]
Here, third cell's rule is $90$, hence depends on left and right neighbors only (see Table~\ref{CArule}). So, the third row of the $T$ is $\begin{bmatrix}
 0 & 1 & 0 & 1
\end{bmatrix}$. Since boundary condition is null, left neighboring cell of first cell and right neighboring cell of last cell are missing.

The global transition function of such a CA can be expressed by $T$. Let us consider $x$ and $y$ be two length-$n$ configurations of an $n$-cell linear CA, and $x$ be the successor of $y$. Then, $y=T\cdot x$, where $x$ and $y$ are considered as vectors.
The matrix $T$, however, can be used to efficiently study the reversibility properties, convergence etc. of a linear (uniform/non-uniform) CA. We briefly discuss this study in Section~\ref{scn_lnraddCA}.

Originally, the matrix $T$ is defined for 3-neighborhood binary CAs. Latter, \cite{vlsi00d,biplab} have extended this tool to represent 3-neighborhood, non-binary, finite CAs. However, the authors have assumed that the states of the cells are the elements of GF($2^p$). These CAs are also linear, and named as {\em hierarchical CAs}.  Further, CAs over GF($2^{p^{q^{r^{\cdots}}}}$) are also designed by \cite{biplabtcad}. 

Apart from 3-neighborhood dependency, 5-neighbor dependent linear CAs are also studied by matrix algebra. For example, \cite{marti2011reversibility} have studied reversibility of such binary CAs with null boundary condition. Whereas, \cite{zubeyir11} have studied reversibility of linear CAs with periodic boundary conditions over $\mathbb{Z}_p$, where $p \geq 2$ is a prime number. For $2$-dimensional linear CAs also, a list of works are reported using matrix algebra as characterization tool, see \cite{Chattopadhyay2d,Siap2d,uguz2013reversibility,chr2D}.

\subsection{Reachability tree} 
\label{tree}
Like matrix algebra, the reachability tree has been proposed to study 1-D non-uniform finite CAs that use only ECAs rules in their rule vectors. Unlike matrix algebra, reachability tree can represent non-uniform non-linear CAs. \cite{Acri04} have first used this tool to study non-uniform finite CAs under null boundary condition. In this section, we first define reachability tree for an $n$-cell non-uniform CA with rule vector ${\mathcal{R}}=\langle {\mathcal R}_0, {\mathcal R}_1, \cdots , {\mathcal R}_{n-1}\rangle$ under null boundary condition, where each rule ${\mathcal R}_i$, $i\in \{0, 1, \cdots, n-1\}$ is an ECA rule.

Recall that an ECA rule can be considered as a collection of RMTs (see Definition~\ref{Def:RMT}) along with their next state values. Let us denote the set of RMTs of ${\mathcal R}_i$ as $Z_8^{i}$. That is, $Z_8^{i}=\{0, 1, 2, 3, 4, 5, 6, 7\}$. However, under null boundary condition, RMTs 4, 5, 6 and 7 of ${\mathcal R}_0$ (similarly, RMTs 1, 3, 5 and 7 of ${\mathcal R}_{n-1}$) are {\em invalid}. So, $Z_8^{0}=\{0, 1, 2, 3\}$, and $Z_8^{n-1}=\{0, 2, 4, 6\}$ for null boundary condition.



\begin{definition}
\label{Def:RTNull}
The reachability tree of an $n$-cell CA with rule vector $\langle {\mathcal R}_0, {\mathcal R}_1, \cdots , {\mathcal R}_i, \cdots, {\mathcal R}_{n-1}\rangle$ under null boundary condition is a rooted and edge-labeled binary tree with $n+1$ levels, where $E_{i.2j} = (N_{i.j}, N_{i+1.2j}, l_{i.2j})$ and $E_{i.2j+1} = (N_{i.j}, N_{i+1.2j+1}, l_{i.2j+1})$ are the edges between nodes $N_{i.j}\subseteq Z_8^{i}$ and $N_{i+1.2j}\subseteq Z_8^{i+1}$ with label $l_{i.2j}\subseteq N_{i.j}$, and between nodes $N_{i.j}$ and $N_{i+1.2j+1}\subseteq Z_8^{i+1}$ with label $l_{i.2j+1}\subseteq N_{i.j}$ respectively  $(0\le i\le n-1$, $0\le j\le 2^i-1)$. Following are the relations which exist in the tree:

\begin{enumerate}

\item \label{rootAtDefNull} {[For root]} $N_{0.0} = Z_8^0 = \{0, 1, 2, 3\}$.

\item \label{edgeAtDefNull} $\forall r\in N_{i.j}$, RMT $r$ of ${\mathcal R}_{i}$  is in $l_{i.2j}$ (resp. $l_{i.2j+1}$), if ${\mathcal R}_{i}[r]$ = 0 (resp. 1). That means,  $l_{i.2j}\cup l_{i.2j+1} = N_{i.j}$ ($0\le i\le n-1$, $0\le j\le 2^i-1 $).

\item \label{nodeAtNodeNull} $\forall r \in l_{i.j}$, RMTs $2r \pmod{8}$ and $2r+1 \pmod{8}$ of ${\mathcal R}_{i+1}$ are in $N_{i+1.j}$ ($0\le i\le n-3$, $0\le j\le 2^{i+1}-1$).

\item \label{n-1AtNodeNull} {[For level $n-1$]} $\forall r \in l_{n-2.j}$, RMT $2r \pmod{8}$ of ${\mathcal R}_{n-1}$ is in $N_{i+1.j}$ ($0\le j\le 2^{n-1}-1$).
	
\item \label{nAtNodeNull} {[For level $n$]} $N_{n,j}=\emptyset$, for any $j$, $0\le j\le 2^n-1$.

\end{enumerate}
\end{definition}

Informally, we can state the reachability tree construction in the following way: Form the root with RMTs 0, 1, 2 and 3 of $\mathcal{R}_0$. Out of the 4 RMTs, the RMTs having next state value 0 are in the label of left edge of the root, and the rest RMTs are in the right edge. Now we can get the nodes of level 1: if an RMT $r$ is in a label of an edge, then put the RMTs $2r\pmod 8$ and $2r+1 \pmod 8$ of $\mathcal{R}_1$ in the node which is connected by the edge. Similarly, we can get next edges and nodes of level 2, and of level 3, etc. However, for the last rule, that is $\mathcal{R}_{n-1}$, all RMTs are not present in null boundary condition. So, only even RMTs of $\mathcal{R}_{n-1}$ can be in the nodes of level $n-1$. Finally, we get the leaves, which are empty.

\begin{figure}
\centering
\includegraphics[height=2.0in, width=4.7in]{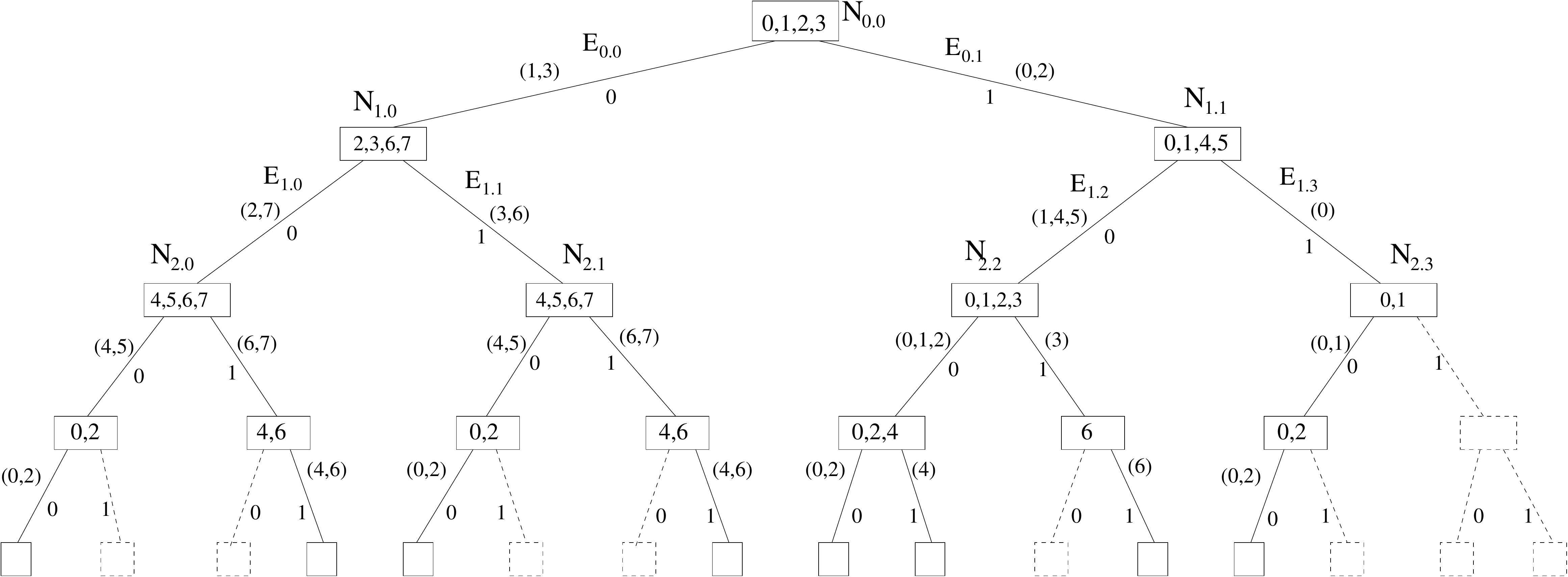}
\caption{Reachability Tree for null boundary $4$-cell non-uniform CA with rule vector $\mathcal{R}=\langle5, 73, 200 ,80\rangle$}
\label{fig: tree}
\end{figure}

Fig.~\ref{fig: tree} shows an example of reachability tree of a 4-cell non-uniform CA with rule vector $\mathcal{R}=\langle5, 73, 200 ,80\rangle$ under null boundary condition. The root is a set \{0, 1, 2, 3\}, and the leaves are empty. It can be observed in the figure that the edges between the nodes of two consecutive levels are formed after applying the corresponding rule on the RMTs. However, an edge $E_{i.2j}$ (resp. $E_{i.2j+1}$) is called as {\em 0-edge} (resp. {\em 1-edge}). In Fig.~\ref{fig: tree}, 0-edges and 1-edges are marked. 

Reachability tree represents {\em reachable configurations} of a CA. A configuration is {\em reachable} if it has at least one predecessor; otherwise it is a {\em non-reachable} (or Garden-of-Eden) configuration. A sequence of edges from root to a leaf represents a reachable configuration when 0-edges and 1-edges are replaced by 0 and 1 respectively. For example, the configuration 0011 of the CA of Fig.~\ref{fig: tree} is reachable. In Fig.~\ref{fig: tree}, some edges (and nodes) are shown by dotted lines. These edges do not exist in the tree. So, a configuration 1110, for example, is non-reachable.

Reachability tree has been utilized to discover many properties of above class of non-uniform CAs, such as reversibility, convergence, etc., see for example \cite{SukantaTH,Adak2016OnSO}. We discuss some of these discoveries in Section~\ref{scn_nlCA}.

When boundary condition changes, the structure of reachability tree also changes. \cite{entcs/DasS09} have first developed this structure to study reversibility of non-uniform CAs under periodic boundary condition. Recently, reachability tree is extended by \cite{jca2015} to study 1-D non-binary finite CAs.


\subsection{$Z$-Parameter, $\lambda$-Parameter, $\Theta$-Parameter}
\label{Sec:Param}
There are some parameters that can be used to characterize some aspects of one-dimensional CAs, such as the $\lambda$ parameter, the $Z$ parameter, and the obstruction ($\Theta$) parameter. The $\lambda$, $Z$, and $\Theta$ parameters are proposed respectively by \cite{langton90}, \cite{WuenscheRePEc,WuenscheI} and \cite{voorhees1997some}. For a $d$-state CA with radius $r$, if $m$ out of the total $d^{2r + 1}$ neighborhood configurations map to a non-quiescent state, then $\lambda$ is defined as: $\lambda = \frac{m}{d^{2r + 1}}$, that is, the percentage of all the entries in a rule table which maps to non-quiescent states. This parameter can be compared with temperature in statistical physics, or the degree of non-linearity in dynamical systems, although these are not equivalent (\cite{LangtonII}).

To track behavior of binary CAs, $Z$ parameter is proposed as an alternate. It takes into account the allocation of rule table values to the sub-categories of related neighborhoods and predicts the convergence of global behavior, extremes of local behavior between order and chaos, surjectivity etc. From a given partial pre-image of a CA state, values of the successive cells can be deduced using this parameter. Two probabilities -- $Z_{left}$ and $Z_{right}$ of the next unknown cell being deterministic are calculated from the rule-table. The $Z$ parameter is the greater of these values and varies between $0$ \& $1$. A derivation of the $Z$ parameter in terms of rule table entries is also given in \cite{WuenscheI}. High value of $Z$ implies that number of pre-images of an arbitrary CA state is relatively small.

 In \cite{voorhees1997some}, the $\Theta$ parameter is defined and shown to characterize the degree of non-additivity of a binary CA rule. It is shown that the $\lambda$ parameter and $\Theta$ parameter are equal respectively to the area and volume under certain graphs. 
These parameters prove their utilization in classification of one-dimensional binary CAs.

\subsection{Space-time diagram and Transition diagram}
Although neither of space-time diagram and transition diagram (or state-transition diagram) is a characterization tool, but these two diagrams have been used to observe and predict the behavior and dynamics of a CA. These diagrams are finite in size. So, for the CAs with infinite or big sizes, only a part of the whole systems can be viewed through these diagrams.

\begin{figure}[hbtp]
     \subfloat[Space-time diagram \label{statespace1}]{%
            \includegraphics[width=0.38\textwidth, height = 5.8cm]{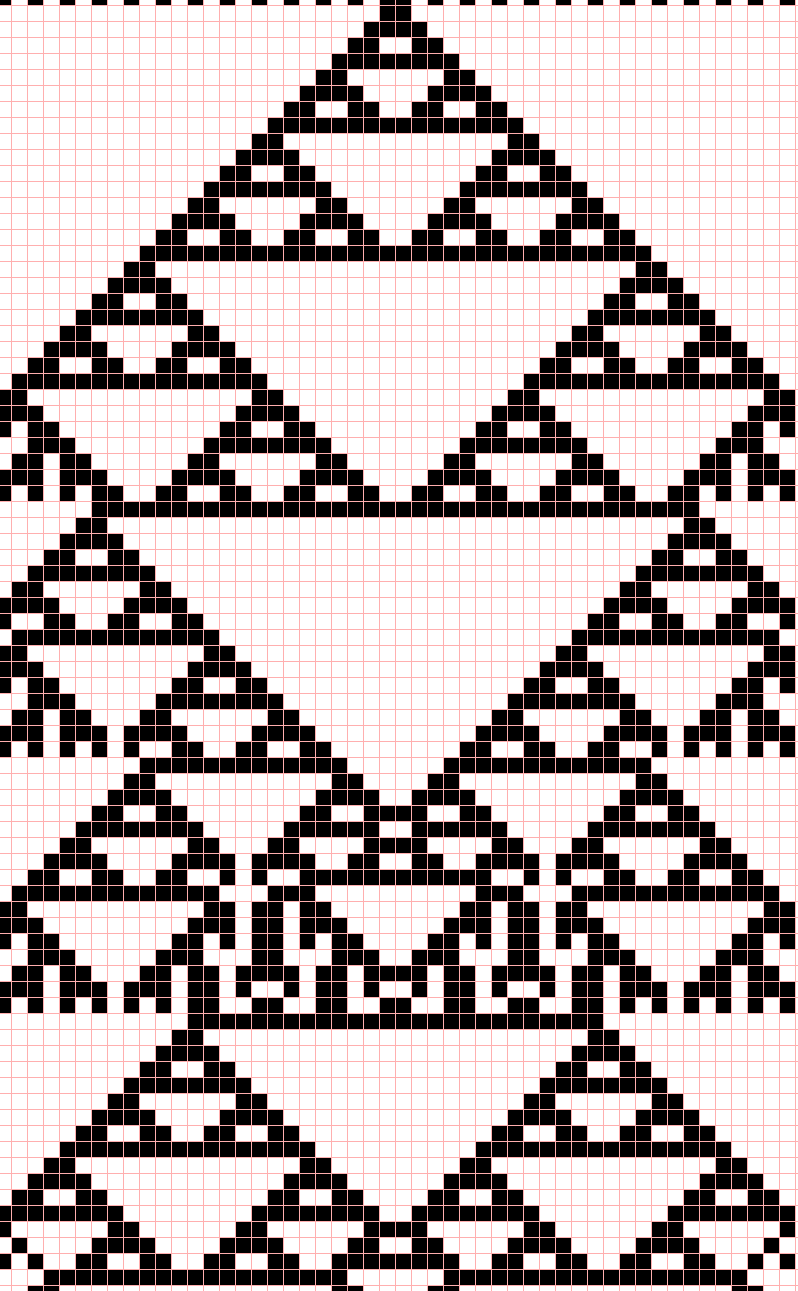}
          }
          \hfill
          \subfloat[Transition diagram \label{transition}]{%
            \includegraphics[width=0.52\textwidth, height = 4.9cm]{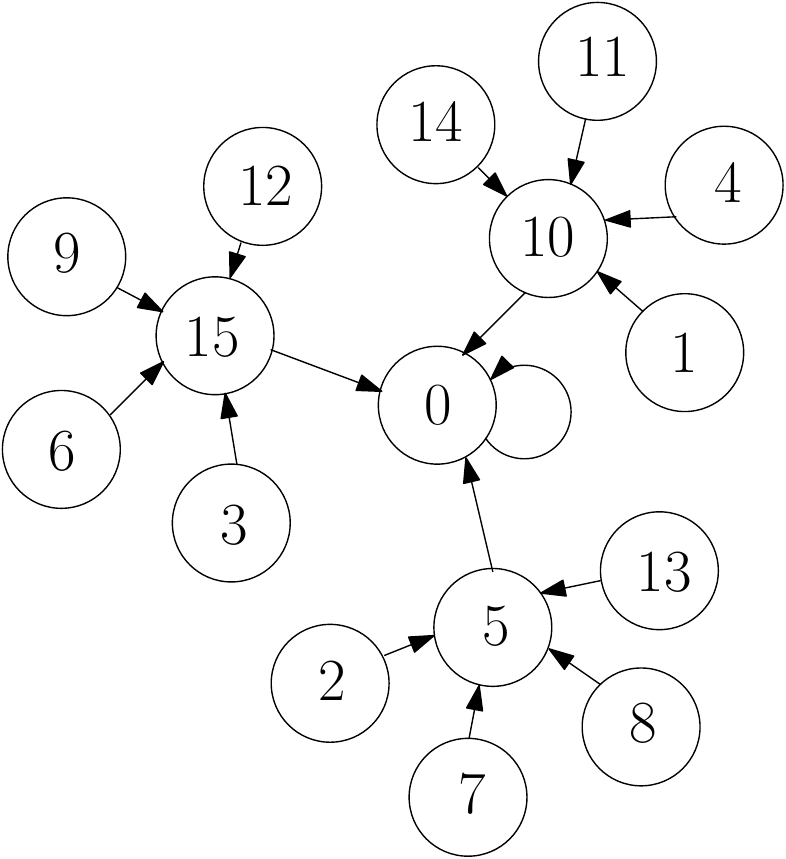}
          }
     \caption{Space-time diagram and transition diagram of $n$-cell ECA $90$ ($2^{nd}$ row of Table~\ref{rt2}) under periodic boundary condition. For (a), cell length $n = 51$, and black implies state $1$ and white implies state $0$. For (b), $n=4$, and the configurations are in their decimal form.}
     \label{fig:evolution diagram}
 \end{figure}

Space-time diagram is the graphical representation of the configurations of a CA at each time $t$. Here, the configuration lies on $x$-axis and $y$-axis represents time. Each of the CA states are generally depicted by some color (see Fig.~\ref{statespace1}). So, the evolution of the CA can be visible from the patterns generated in the state-space diagram. This has been used to study the nature of CA in a set of papers, see for example \cite{Wolframbook1,DennunzioFP14}. Some packages are available in public domain which can be used to observe space-time diagrams of CAs - \emph{Fiatlux} (see \cite{FiatNazim}) is one of them.

Transition diagram, also called as state transition diagram, of a CA is a directed graph whose vertices are the configurations of the CA, and an edge from a configuration $y$ to another configuration $x$ represents that $x$ is the successor of $y$ (see Fig.~\ref{transition}). Sometime the directions are omitted if they can be understood from the context. Transition diagrams are heavily used in the works of \cite{ppc1}. It can be noted that, the dynamic behavior of any CA can also be visualized and studied using its transition diagram. \emph{Discrete Dynamics Lab}, an online laboratory (\cite{DDLab}), is a good place to get transition diagrams of CAs.

\section{Global behavior of Cellular Automata}
\label{sscn_prpCA}
The most exciting aspect of CAs is their complex global behavior, which is resulted from simple local interaction and computation.
The CAs have many elementary properties of the real world, such as reversibility and conservation laws, chaotic behavior, etc.
These properties motivate researchers to simulate physical and biological systems by CAs. \cite{Vichn84} has targeted to simulate physics by CAs. Lattice gases (\cite{Frisc86}), Ising spin models (\cite{Grins85}), traffic systems (\cite{379-440-26,PhysRevLett-35}) etc. are also simulated using CAs. The book of \cite{Chopard} shows about these works. Apart from the ability of modelling physical world, the CAs are capable of performing several computational tasks. In this section, we survey some of the global behaviors of CAs such as universality, reversibility, conservation laws, computability, etc.

\subsection{Universality}
The first feature that has attracted the researchers is the capability of CAs in performing universal computation. An arbitrary Turing machine (TM) can be simulated by a CA, so universal computation is possible by CA. The universality of CAs has been first studied by \cite{Thatcher,Arbib66,banks1970universality}. However, the idea of simulating TM does not use the parallelism property of a CA. In \cite{Smith:1971}, the existence of computation-universal cellular spaces with small neighbor-state product is proved.
Later, \cite{Morita89,Dubacq95} have shown the procedure of simulating a TM by a reversible CA. 
Computation universality of one-way CAs and totalistic CAs are also reported by \cite{iirgen1987simple}. Computational universality is also shown following very simple rules, such as Game of Life (\cite{Gardner71}), the billiard ball computer (block CAs) (\cite{Durand-Lose98}). The existence of computation-universal one-dimensional CA with $7$-states per cell for radius $r=1$ and $4$-states per cell for $r=2$ is proved in \cite{lindgren90}. A very simple automaton, ECA 110 of \cite{wolfram84b} is shown as computationally universal by \cite{cook2004universality}. 

Apart from simulating a Turing machine, a CA can simulate another CA. If a CA can simulate all CAs of the same dimension, it is called intrinsically universal (\cite{Kari05}).
 The smallest intrinsically universal CA in one-dimension is reported in \cite{ollinger2003intrinsic}. For two dimension, nonetheless, the same is a $2$-state and $5$-neighborhood CA (\cite{banks1970universality}). Some other notable related works are \cite{Martin94,ollinger2002quest}.

\subsection{Invertibility and reversibility}
For long, the questions of invertibility have been the major focus of research in CAs. If the global transition function $G$ of a CA is one-to-one, the CA is termed as an {\em injective} CA. However, it is called \emph{surjective} if the function is onto and bijective if $G$ is both onto and one-to-one.
The study of reversibility of CA was started with \cite{hedlund69} and \cite{Richa72}. From \cite{hedlund69}, we get that a function $G:\mathcal{S}^{\mathbb{Z}^D}\rightarrow \mathcal{S}^{\mathbb{Z}^D}$ is the global transition function of a $D$-dimensional (infinite) CA if and only if $G$ is continuous, and it is shift-equivalent (this result is known as {\em Curtis - Hedlund - Lyndon} Theorem). As a consequence of this result, we get that a CA $G$ is {\em reversible} if and only if it is a bijective map. In fact, Hedlund and Richardson independently proved that all injective CAs are reversible. 

In their seminal paper, \cite{Amoroso72} have shown an effective way to decide reversibility of $1$-dimensional infinite CA, on the basis of the local rule. In \cite{di1975reversibility}, a decision algorithm is reported for CAs with finite configurations.
An elegant scheme based on de Bruijn graph for deciding the reversibility of a one dimensional CA is presented in \cite{suttner91}. However, \cite{Kari90} has shown that no algorithm can decide whether or not an arbitrary 2-D CA is reversible. This result can also be extended to higher dimensional CAs. An interesting result has been reported by \cite{toffoli77}, which says, a $(D+1)$-dimensional reversible CA can simulate any $D$-dimensional CA. 
Later, \cite{Morita89,morita1995reversible} have shown that a $1$-D reversible CA can simulate reversible Turing machines as well as any $1$-D CA with finite configurations. Since reversible Turing machines can be computationally universal, we can find universal 1-D reversible CA. Some other notable works on reversible CAs are \cite{Maruoka197947,culik1987invertible,Durand93,Kari94,Kari2005,moraal2000graph,MoraMM06,Soto2008,Morita2008101}.

However, all the works reported above deal with infinite CAs. In case of finite CAs, things are bit different. For example, the algorithms of \cite{Amoroso72} can not successfully work on finite CAs. 
\cite{marti2011reversibility,zubeyir11} have studied reversibility of 1-D linear finite CAs having only two states.
Recently, \cite{jca2015} have reported an algorithm to decide the reversibility of 1-D finite CAs. The reversibility of 1-D non-uniform CAs has also deeply studied. We discuss about these studies in Section~\ref{scn_eca}.

\subsection{Garden-of-Eden} 
One of the earliest discovered results on CAs was the \textit{Garden-of-Eden} theorems by \cite{moore1962machine} and \cite{Myhill63}.
\textit{Injectivity} and \textit{surjectivity} properties of CA are correlated by these theorems. A configuration is named as a Garden-of-Eden configuration, if it does not have a predecessor; that is, if it is a {\em non-reachable} configuration. In 1962, Moore has shown that the existence of mutually erasable configurations in a two-dimensional CA is sufficient for the existence of Garden-of-Eden configurations (\cite{moore1962machine}). He has also claimed that existence of a configuration with more than one predecessors ensures existence of another configuration without any predecessor.
However, in \cite{Myhill63}, the reverse was proved. He has shown that the extant of mutually indistinguishable configurations is both necessary and sufficient for the extant of Garden-of-Eden configurations.  

In \cite{amoroso1970garden}, the equivalence between the existence of mutually erasable configurations and mutually indistinguishable configurations has been established. This implies that the converse of Moore's result is true as well. It has also been shown in the paper that, for finite configurations (Definition~\ref{Def:FiniteConf}) both of the above conditions remain sufficient, but neither is then necessary. For finite configurations, a CA is irreversible if and only if a Garden-of-Eden configuration exists. And, a CA is surjective, if and only if it is bijective (\cite{amoroso1970garden}). So, the existence of Garden-of Eden configurations violates the injectivity property. The relation between global function of infinite CA and its restriction to finite configurations has been established in \cite{Richa72}. Some other important results are listed in \cite{maruoka1976condition,sato77,toffoli90,Kari05}.

Garden-of-Eden theorems for CAs have further been extended to Cayley graphs of groups by \cite{doi:10.1137/0406004,ceccherini1999amenable}. In \cite{capobianco2007surjectivity}, the surjectivity and surjunctivity of CA in Besicovitch topology is recoded. 
Moreover, in \cite{margenstern1999polynomial,Margenstern200199}, CA is defined in hyperbolic plane. In \cite{margenstern2009garden}, it is depicted that, the injectivity and surjectivity properties, proved by Moore and Myhill, are no longer valid for CAs in the hyperbolic plane.

\subsection{Topology, dynamics and chaotic behavior of CAs}
The global transition function $G$ of an infinite CA is a continuous map on the compact metric space $\mathcal{S}^{\mathbb{Z}^D}$ according to the topology induced by the metric $d$. This metric $d$ is usually considered as the Cantor distance. Therefore, a CA can be viewed as a discrete time dynamical system $\langle \mathcal{S}^{\mathbb{Z}^D}, G\rangle$. So, the (infinite) CAs can be studied by topological dynamics and chaos theory.

Many definitions of chaos, however, use the notion of sensitivity to initial condition. A CA is {\em sensitive to initial condition}, or simply {\em sensitive} if and only if there exists a $\delta >0$ such that, $\forall x \in {\mathcal{S}^{\mathbb{Z}^D}} ~~ \forall \epsilon >0 ~~ \exists y \in \mathcal{S}^{\mathbb{Z}^D} ~~ \exists n \in \mathbb{N}: ~~~ d(x,y) < \epsilon~~  \mbox{and}~~ d(G^n(x),G^n(y))\geq \delta$. That is, in a sensitive CA, a (small) change in initial condition would greatly affect the CA in future. A popular definition of chaos is due to \cite{Devaney}, which says that a dynamical system is chaotic if and only if it is transitive, has dense periodic points, and sensitive to initial conditions. However, a CA is {\em transitive} if and only if for any two non-empty open subsets $U$ and $V$ of $\mathcal{S}^{\mathbb{Z}^D}$, there exists an $n\in\mathbb{N}$ so that $G^n(U)\cap V \neq \emptyset$. Further, a CA has {\em dense periodic points} if and only if the set $\{x\in \mathcal{S}^{\mathbb{Z}^D}|\exists k\in\mathbb{N}: G^k(x)=x\}$ of all periodic points is a dense subset of $\mathcal{S}^{\mathbb{Z}^D}$. It has been proved by \cite{CM96} that for CAs, transitivity implies sensitivity to initial condition. So, a CA is chaotic in Devaney's sense if and only if it is transitive and has dense periodic points.

Apart from the Devaney's definition, there are more restrictive definitions of chaos, such as Knudsen chaos, positively expansive chaos, etc. A CA is {\em chaotic} according to the definition of Knudsen if and only if it has a dense orbit and sensitive to initial conditions. The CA $G$ has a {\em dense orbit} if and only if there exists a configuration $x$ so that $\forall y\in \mathcal{S}^{\mathbb{Z}^D} ~~ \forall\epsilon >0~~ \exists n\in\mathbb{N}: ~~ d(G^n(x),y)<\epsilon$. On the other hand, a CA is {\em positively expansive chaotic} if and only if it is transitive, has dense periodic points, and positively expansive. Positive expansivity is a stronger form of sensitivity. A CA is {\em positively expansive} CA if and only if there exists a $\delta>0$, such that for any two different configurations $x$ and $y$, $d(G^n(x), G^n(y))\geq \delta$ for some $n\in\mathbb{N}$.  However, these definitions of chaos are related. The chaotic behavior of CAs are studied by a number of works; some examples are \cite{DCTMitchell93,Margara99,Cattaneocht,AcerbiDF09}.

Based on their dynamic behavior, CAs are classified into different classes. One such classification, depending on their degree of {\em equicontinuity}, is due to \cite{Kurka97}. A configuration $x\in \mathcal{S}^{\mathbb{Z}^D}$ is an {\em equicontinuity} point of the CA if for every $\epsilon>0$, there is a $\delta >0$ such that $\forall y\in \mathcal{S}^{\mathbb{Z}^D}: d(x,y) <\delta ~~\mbox{implies}~~ d(G^n(x),G^n(y))<\epsilon$ for all $n\in\mathbb{N}$.
Now, a CA is called equicontinuous if all configurations are equicontinuity points. There is a connection between sensitivity and equicontinuity points of a CA: a CA $G$ is sensitive to initial conditions, if the CA does not have any equicontinuity points. Following is the classification of Kurka:
\begin{enumerate}
\item[(1)] equicontinuous CAs,
\item[(2)] CAs with some equicontinuity points, 
\item[(3)] sensitive but not positively expansive CAs, and
\item[(4)] positively expansive CAs.
\end{enumerate}
\cite{durand2003undecidability} have studied the decision problems to determine whether a given CA belongs to a given class, and they have shown that most of the problems are undecidable. However, for the additive CAs, \cite{DENNUNZIO20094823} have studied the directional dynamics, and then classified them.
 
Apart from the above work, there are many other works, specially for the 1-D case, that target to classify CAs depending on their dynamical behavior. \cite{wolfram84b} has reported a classification of ECAs without considering a precise mathematical definition. \cite{culik88} have formalized Wolfram's classification. In fact, different parameters, discussed in Section~\ref{Sec:Param}, have been developed to classify the CAs depending on their chaotic behavior. For the two-dimensional space, the generalization of the parameters -- sensitivity, neighborhood dominance and activity propagation, is reported in \cite{deOliveira20061}.

Another studied behavior of CAs is {\em nilpotency}. A CA is {\em nilpotent} if for each configuration $x$, $G^n(x)=c$ is a singleton set for sufficiently large $n$. Obviously, $c$ is a fixed point. \cite{Culik90} have shown that for two or more dimension, the nilpotency of CAs is undecidable. The same result for $1$-D CAs has been proved by \cite{Kari92a}. 
Some more works to study the dynamical properties of CAs, using expansivity, subshift, homomorphism automorphisms and endomorphisms are reported in \cite{blanchard1997dynamical,WARD1994495}. \cite{Sheresh} has defined the left and right Lyapunov exponents for $1$-dimensional CAs. \cite{Finelli199} generalized the theory of Lyapunov exponents for $D$-dimensional CAs and proved that all expansive CAs have positive Lyapunov exponents for almost all the phase space configurations.

\subsection{Randomness}

Stephen \cite{Wolfram85c} has introduced CAs as an excellent source of pseudo-randomness. Massive parallelism, simplicity and locality of interactions of CAs, offer many benefits over other techniques, specially in case of hardware implementation. These benefits along with pseudo-randomness have made CAs as an area of extensive research in VLSI circuit testing (see for example \cite{Horte89a,ppc1,tcad/DasS10}), Monte-Carlo simulations (see \cite{870571}), Field Programmable Gate Arrays (see as example \cite{comer2012random}), cryptography (see as example \cite{Wolfr86b,DBLP:journals/ccds/DasC13,DBLP:journals/jca/DasR11,Formenti2014,Leporati2014CryptographicPO}) etc. In fact, the most appealing application of pseudo-randomness of CAs is in the domain of cryptography. We briefly discuss these applications in Section~\ref{scn_appCA}.
However, most of the works on pseudo-randomness have been divided in mainly two directions:

-- Most of the research have been going on in the first direction, to generate pseudo-random \emph{numbers} using CAs. Here, generally an integer $X_i$ is generated between zero and some number $m$ (word size of the computer), where the fraction $U_i = \frac{X_i}{m}$ is the real number, uniformly distributed between $0$ and $1$. This number can be generated in several ways. For example, in \cite{wolfram86c}, the sequence is generated by ECA $30$ from the single cell with initial state $1$ among all cells, initiated with state $0$. In a recent work, \cite{IJMC2017} have used a 3-state 3-neighborhood 1-D CA, and considered a small window of cells. The base-3 numbers, observed through the window, are considered the pseudo-random numbers. Sometime whole configurations of a finite CA are also considered as numbers. Some other works of generating pseudo-random numbers are \cite{Marco00,alonso2009elementary}. In \cite{Tomassini96,wang2008generating}, some optimization algorithms are applied to CAs, whereas in \cite{Guan03,Guan04} dynamic behavior is allowed in the cells to generate the pseudo-random numbers.

-- In the second direction, pseudo-random \emph{patterns} are generated using finite CAs. Here, pattern means the configuration of a CA of length $n$, where each cell can take any of the CA states. Note that, in a pattern, individual cell values have significance. For example, a sequence $\langle0110\rangle$ can be a treated as a number $6$, but in case of pattern, it is ``$0110$''. Some notable works on pseudo-random pattern generation using non-uniform CAs are in \cite{ats03,SukantaTH}. Here finding of the minimum cell length is important. For example, in \cite{SukantaTH}, a $45$-cell CA based pseudo-random pattern generator (PRPG) is designed which beats all existing PRPGs.

\subsection{Conservation law}
Apart from reversibility, there exist other conservation laws, which are equally important in physics. Various direction of conservation (invariants) laws in CAs are reported in \cite{fredkin82,pivato2002-45,Boccara02}. Among them number conserving CAs (NCCAs) are the most studied and used concept. Let us interpret the states of cells as numbers. Then, a CA is an NCCA if the sum of the states remains invariant during evolution of the CA. NCCAs are defined with respect to the spatially periodic configurations and finite configurations. Boccara and Fuk$\acute{s}$ have given the necessary and sufficient conditions of a $1$-D CA to be NCCA, initially for $2$ states per cell in \cite{boccara-1998-31} and then for any arbitrary number of states in \cite{Boccara02}. In \cite{pivato2002-45}, a general treatment to conserved quantities in $1$-D CA is reported. Fuk$\acute{s}$ has shown that motion representations can be constructed by $1$-D binary NCCAs. \cite{pCA18} has also extended the work to probabilistic CAs. NCCAs in higher dimensions are also studied by \cite{Durand03}. Further, universality and other dynamics of NCCAs are explored in \cite{Moreira2003711,Formenti2003269}. 

NCCAs have widely appeared as the models of highway traffic. The notable works in this regard are \cite{nagel1992cellular,PhysRevE-40,PhysRevLett-35,379-440-26}. \cite{Kohyama01011989-27,-28} has used NCCAs for particle conservation. In $2$-dimension, the Margolus CA is a number conserving CA; see \cite{Margolus198481}. Apart from NCCAs, additive conserved quantities in CAs are introduced and investigated in \cite{Hattori}. In \cite{takesue1995staggered}, a necessary and sufficient condition for a given CA rule to profess a staggered invariant is studied. In \cite{Morita98,Morita:99}, the computational universality of partitioned number-conserving (and reversible) CA is shown by simulating a universal counter machine. However, these CAs are not exactly the same as NCCAs, because reducing a partitioned CA to a non-partitioned one is not number preserving. The concept of extending conserved quantities to that of monotone quantities was presented in \cite{Kurka:2003}, considering CAs with vanishing particles. Number conserving property of CAs has also been studied by applying the communication complexity approach; see for example \cite{GOLES20113616}.

In \cite{das2011characterization}, non-uniform NCCAs are characterized. This paper has reported $O(n)$ time algorithms for verification and synthesis of an $n$-cell non-uniform NCCA. \cite{DennunzioFP14} have given a generalized definition of number conservation of non-uniform CAs. Further, number conservation property of ECA under asynchronous update has been studied in \cite{hazari14}.

\subsection{Computational tasks}

In the CAs literature, two domains related to computation are mostly studied:
density classification task and synchronization problems. The problem statement of density classification can be summarized as follows - any initial configuration having more $1$s ($0$s) than $0$s ($1$s) must converge to all $1$s ($0$s) configuration. However, \cite{PhysRevLett.74.5148} have proved that it is impossible to solve this problem with $100\%$ accuracy. Because of the impossibility of solving the standard density classification task, research efforts have shifted towards finding the best rule which can solve the problem {\em almost} perfectly -- for one-dimensional CA by \cite{Kari:2012:MTC:2385073.2385086}, for two-dimensional CA by \cite{morales01,deOliveira20061}, for non-uniform CA by \cite{DCTMaiti06} and stochastic CA by \cite{fates00608485}. However, \cite{Fuk05} has shown that the density classification task is solvable by running in sequence the trivial combination of elementary rules $184$ and $232$. A good survey about the problem can be found in \cite{Oliveira13}.

The synchronization problems, like \textit{firing squad}, \textit{queen bee}, \textit{firing mob}, etc. are studied by CAs. The goal of firing squad synchronization problem is designing a CA, which initially has one active cell and evolves to a state, where every cell is simultaneously active. A good number of works on this problem are found in literature; some examples are \cite{FSSPCA2,FSSPCA3,MANZONI2014108}. The firing mob problem, which is a generalization of firing squad problem, is solved by \cite{Culik91}.
However, the queen bee and leader election problems are considered as the inverse problem of firing squad problem. Here, initially states of all cells are identical and by choosing a proper rule, a cell comes to a special state. \cite{Smith76} has first introduced the problem. Latter the problem has been explored by a number of researchers, such as \cite{Mazoyer,BeckersW01,Stratmann154,banda1}.

Another computation problem, named early bird problem has been defined and investigated first by \cite{rosen}. Here, any cell in the quiescent state may be excited by the outside world. These excitations result in special ``bird'' states instead of the quiescent states. The task is to give a transition function such that, after a certain time the first excitation(s) can be distinguished from the latter ones. This problem has been studied by \cite{Vollmar2,Legendi,Legendi1}. A variation of this problem is the distributed mutual exclusion problem, where some cells of initial configuration are in a special state (called {\em critical section requesting state}) and during evolution of the CA, these cells will be in another special state (called {\em critical section executing state}) one-by-one. Recently this problem is explored by \cite{SRoy2017}. Some other computation problems, such as French flag problem (\cite{Herman01081973}), shortest path problem (\cite{short12}), generating discretized circles and parabolae in real time (\cite{Delorme1999347}) etc. are also discussed in literature.

\section{Non-uniformity in cellular automata}
\label{scn_nunCA}

Conventionally, all the variants of CAs possess basic three properties - \emph{uniformity, synchronicity} and \emph{locality}. The uniformity refers to that each of the CA cells are updated by the identical local rule. The synchronicity implies that all the cells are updated simultaneously; whereas locality refers to that the rules act locally and neighborhood dependencies of each cell is uniform. Note that, the cells perform computation locally, and the global behavior of CA is received due to this local computation only. However, synchronicity is a special type of uniformity, where all the cells are updated simultaneously and uniformly. In fact, uniformity is everywhere in CA, in local rule, cell update and in lattice structure. We can summarize this in the following way:
 \begin{itemize}
 \item Uniformity in update: all cells are updated simultaneously in each discrete time step.
 \item Uniformity in lattice structure and neighborhood dependency:  lattice structure is uniform and each cell follows similar neighborhood dependency.
 \item Uniformity in local rule: each of the cells updates its state following the same rule.
 \end{itemize}
 
Over the years, researchers have successfully been using classical CA as a modelling tool. However, it has become apparent that many phenomena, such as chemical reactions occurring in a living cell, are found in nature which are not uniform. These new modelling requirements led to a new variant of the CAs. As a result, non-uniformity in CAs has been introduced. Following are the main three variants of non-uniformity in CAs, which we get after relaxing above mentioned constraints of uniformity.

\begin{enumerate}
\item Asynchronous cellular automata (ACAs): the cells are not updated at the same (discrete) time step and can be independently updated - breaks the uniform update constraint (discussed in Section~\ref{sscn_ACA}).

\item Automata Network: the CA is on a network and the states of the node evolve with neighborhood defined by the network - breaks uniform neighborhood constraint (discussed in Section~\ref{sscn_NA}).

\item Hybrid or non-uniform cellular automata: cells can assume different local transition functions - breaks uniform local rule constraint (discussed in Section \ref{sscn_HCA}).
\end{enumerate}

\subsection{Asynchronous Cellular Automata ($ACA$s)}
\label{sscn_ACA}
Like other synchronous systems, a CA also assumes a global clock which forces the cells to get updated simultaneously. This assumption of global clock is not very natural, and is relaxed in ACAs. 
The concept of ACAs and their computational ability has first been developed by \cite{naka}, it has further been studied by \cite{GOLZE1978176,Nakamura22,Hem82,Ingerson84,LeC89}. ACAs have been developed on two-dimensional grid by \cite{Cor} to report the concurrent situations emerged in distributed systems.

The word `asynchronism' means that the parts of the system do not share the same time. In asynchronous CAs, cells are independent and so, during the evolution of the system, the cells are updated independently. There are several interpretations on the way of applying asynchronism. By simplifying, it can be said that asynchronism is to break the perfect update scheme. The main asynchronous updating schemes found in literature are \emph{fully asynchronous updating} and \emph{$\alpha$-asynchronous updating}.
\begin{itemize}
\item Under \emph{fully asynchronous updating} scheme, a cell is chosen uniformly and randomly at each time step to update. That is, at each step, only one cell is updated.
\item In \emph{$\alpha$-asynchronous updating} scheme, each cell is updated with probability $\alpha$. This implies that a cell does not apply the rule with probability $1-\alpha $, and stays in its old state.
\end{itemize}
The parameter $\alpha$ is known as synchrony rate. Obviously, when $\alpha =1$, the CA becomes synchronous. In that sense, the classical CAs are special case of $\alpha$-asynchronous CAs. Latter, \cite{probing12} have used other asynchronous updating schemes: $\beta$- and $\gamma$-asynchronism. Then, \cite{Dennunzio13} have developed an $m$-asynchronous CA and generalized the various updating methods used so far. A survey on asynchronous update schemes can be found in \cite{Fates14}.

In one of the pointing work, \cite{Hem82} has shown the computation equivalence of synchronous and asynchronous cellular space. He has also shown that, any $d$-state deterministic rule can be simulated by a $3d^2$ state asynchronous rule with same neighborhood dependency. From a different point of view,  \cite{GOLZE1978176} has shown that, a $(D+1)$-dimensional asynchronous rule can simulate a $D$-dimensional synchronous rule. \cite{Dennunzio16} showed how fully asynchronous CA can simulate universal Turing machine. 

In an early work, \cite{Golze1982121} have shown that ACA can implement Petri net. A Petri net is a directed bipartite graph which is used to model distributed systems. \cite{Cor} have shown latter that ACAs can be used as models of concurrency and distributed systems. The computing abilities of these ACAs have further been investigated by \cite{Pighizzini1994179,Droste20001}. As shown by \cite{MANZONI2014108}, ACAs can be used to address firing squad synchronization problem.

In \cite{PhysRevE}, the change that occurs in the Game of Life, when the sites get updated with a given probability, are identified. \cite{Ruxton} have analyzed the sensitivity of ecological system modelled by simple stochastic cellular automata to spatio-temporal ordering.
In \cite{Tomassini02,fates00608485}, asynchronous rules are used to address the density classification problem. \cite{Suzudo2004185} has studied the usage of genetic algorithms for determining the mass-conservative (also called number-conserving) asynchronous models that would generate nontrivial patterns. 

It is thought that reversibility and asynchronism are two opposite terms. However, the question of reversibility in ACAs has been studied by \cite{sarkar2012reversibility,SethiD14}. Their approach is to study the possibility of returning back to the initial configuration after a finite number of time steps. These so-called reversible ACAs have been used by \cite{ACASir} to generate patterns with specific Hamming distance. These reversible ACAs have further been utilized by \cite{Sethi2016} in symmetric-key cryptography. \cite{Mariot2016} has introduced the notion of asynchrony immunity for CAs that could be used for crytography. In \cite{Manzoni2012}, the dynamical properties of CAs, such as injectivity, surjectivity, permutivity, sensitivity, expansivity, transitivity, dense periodic orbits and equicontinuity are re-defined for the asynchronous CAs.

Convergence of ECAs to fixed points under fully asynchronous update has been studied by \cite{Fates20061}. The convergence time of these ACAs have also been explored in this work. Recently, \cite{Sethi2015,CPLX:CPLX21749} have further studied the convergence of these ACAs, and have used the convergent ACAs in designing an efficient two class pattern classifier.

\subsection{Automata Network}
\label{sscn_NA}

Traditionally, cellular automata consist of a regular network with local uniform neighborhood dependency. However, in automata network (also called as cellular automata network), this uniform local neighborhood dependency is relaxed. Here, cellular automata rules allow a cell to have an arbitrary number of neighbors, and thus can be set to work on any given network topology, as shown, for example, in \cite{Marr20}. As a matter of fact, the rules of automata networks can not be always local. And, since the non-local and local rules are different, it is, therefore, expected that the non-local rules may lead to different behavior from the conventional local rule-based CAs. Some example works in this regard are \cite{boccara1994some,newman99,yang2007}. 

The earliest version of such non-uniformity in neighborhoods are found in \cite{Jump74,Smith76}.
From the 1990s, networks have been used as an important model for solving different complex problems; see as example \cite{Adami199529,watts1998collective}. In fact, after the work of \cite{watts1998collective}, researchers become more interested on automata networks. 
As noted by \cite{Tomassini15}, we get a wider class of generalized automata networks by extending standard lattice cellular automata and random Boolean networks. 
\cite{tomassini29,Darabos7} have shown that automata networks with arbitrary topologies perform better than the regular lattice structures for the majority and synchronization problems. The work of \cite{Cor}, which is a pioneering work on ACAs and which models concurrency and distributed systems, also uses automata network. \cite{yang2007} have developed a new type of small-world cellular automata by combining local updating rules with a probability of long-range short-cuts to simulate the interactions and behavior of a complex system.

\cite{8718492.ch2} have investigated automata network as algebraic structures and developed their theory in line with other algebraic theories, such as semi-groups, groups, rings and fields. They have also shown a new method for the emulation of the behavior of any (synchronous) automata network by the corresponding asynchronous one. \cite{Kayama2011} have shown the network representation of Game of Life, where the characteristics is like one of Wolfram's class IV rules. In \cite{Kayama2012}, the network derived from ECAs and five neighbor totalistic CA rules are further reviewed. However, the studies in this area are still at a very early stage.

\subsection{Non-uniform CAs or Hybrid CAs}
\label{sscn_HCA}

Among the above mentioned models, the most popular and studied model is {\em Hybrid} CA or {\em Non-uniform} CA, where the cells can use different local rules. The study of the non-uniform CA has been started with \cite{Pries86}, where the authors have studied the group properties of $1$-dimensional finite CAs under null and periodic boundary conditions. In that work, a special type of non-uniform CAs have been investigated, where the cells use Wolfram's CAs rules (see Table~\ref{rt2}). Since then, however, the major thrust of non-uniform CAs research has been on this class of CAs, see for example \cite{Horte89c,Das91,ppc1,entcs/DasS09}. We dedicate the next section to survey this class of
non-uniform CAs.

In recent years, the generalized definition of non-uniform CAs has been given by \cite{CattaneoDFP09,DennunzioFP12}, where the cells may follow different rules with different neighborhood dependencies. Formenti and his colleagues have been investigating this class of non-uniform CAs, and they have identified various sub-classes of these CAs. Some basic global properties of non-uniform CAs, such as surjectivity, injectivity, equicontinuity, decidability, structural stability etc. have also been explored by  \cite{CattaneoDFP09,DennunzioFP12,DennunzioFP14,salo2014realization}.

\section{Non-uniform ECAs}
\label{scn_eca}

Nowadays, research on non-uniform CAs has gained a popularity. The researchers have been exploring them from various directions, and proposing generalized definition of non-uniform CAs. However, since late $1980$s until today, the primary focus of non-uniform CAs research has been on a special class of $1$-dimensional CAs, where the cells follow Wolfram's rules. The main reason of choosing this class of CA is two-fold -- $(1)$ Wolfram, in early $1980$s, showed the efficacy of $3$-neighborhood binary CAs in modelling physical systems and in producing complex global behavior, and $(2)$ ease of implementing Wolfram's CAs rules in hardware. We call these CAs as non-uniform ECAs, to differentiate them from others. Since the early days, these CAs are explored targeting some hardware-related problems. Obviously, these CAs are finite. In this section, we discuss only about finite non-uniform ECAs.

\begin{figure}
\centering
\includegraphics[width=4.2in,height=1.5in]{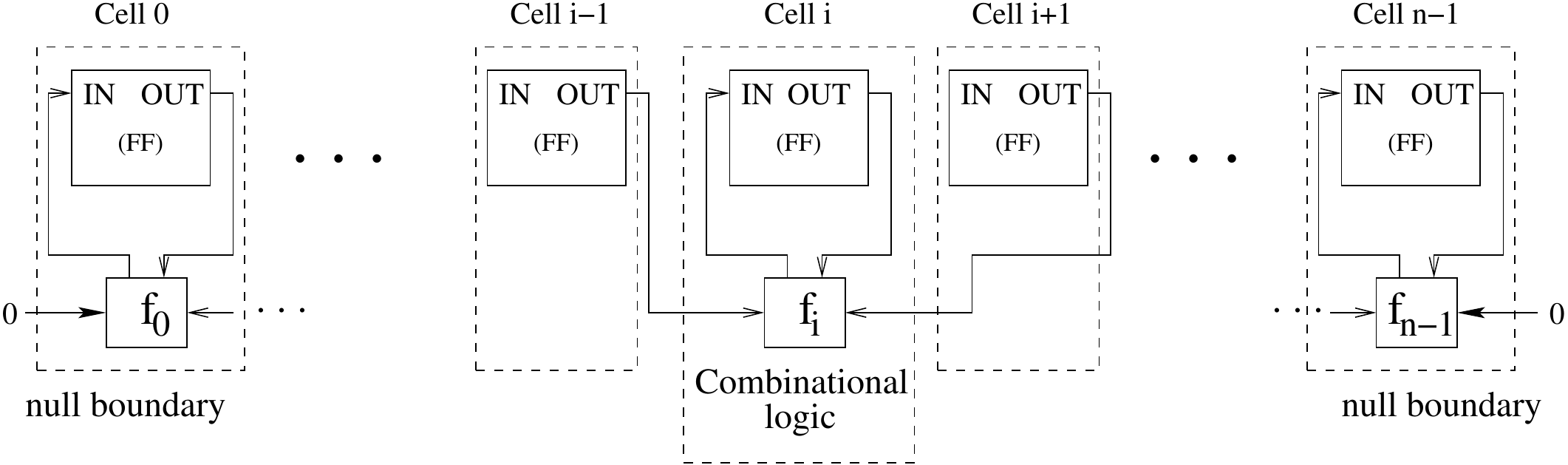}
\caption{Implementation of an $n$-cell non-uniform ECA under null boundary condition}
\label{Fig:ImplNullCA}
\end{figure}

Fig.~\ref{Fig:ImplNullCA} shows hardware implementation of an $n$-cell non-uniform ECA under null boundary condition. Here, each cell consists of a flip-flop (FF) to store the state of a cell and a combinational logic circuit to find the next state of the cell. Due to similarity of this structure with finite state machine, many authors, who have been working with finite non-uniform ECAs, consider these CAs as finite state machine (FSM). As a consequence, these authors call the configurations of CAs as states. 

As we have discussed, however, different cells of a non-uniform ECA may use different rules, so we need a rule vector  $\mathcal{R}=\langle \mathcal{R}_0, \mathcal{R}_1, \cdots \mathcal{R}_{n-1} \rangle$ to define these CAs (see Definition~\ref{Def:RuleVector}). Here, cell $i$ uses rule $\mathcal{R}_i$, and $\mathcal{R}_i$ is presented as a decimal number as shown in Table~\ref{rt2}. Obviously, uniform or classical CAs are special cases of non-uniform CAs where $\mathcal{R}_0=\mathcal{R}_1= \cdots =\mathcal{R}_{n-1}$.
In this section, we will survey the research organized to understand the behavior of non-uniform ECAs, with respect to two categories: additive/linear CAs and non-linear CAs.

\subsection{Linear/additive CAs}
\label{scn_lnraddCA}
A CA is linear if  $G$, the global transition function of the CA, is linear.
There are seven ECAs having rules $60, 90, 102, 150, 170, 204$ and $240$ which satisfy the linearity conditions (we have excluded rule $0$ from the list, which also satisfies the conditions but a trivial one). These are linear ECAs, and the rules are linear rules (see also Section~\ref{Sec:rule}). However, there is no additive ECAs other than these ECAs. So in literature, the terms ``linear CA'' and ``additive CA'' are used interchangeably.

A (finite) non-uniform ECA is linear/additive if and only if each of the rules in rule vector is linear/additive. Since there are 7 linear/additive rules, the rule vector $\mathcal{R}$ of a linear/additive non-uniform ECA is to be designed with only these 7 rules. The linear ECAs rules can be expressed by $XOR$ logic (see Table~\ref{CArule}, which is reproduced directly from \cite{ppc1}). A linear/additive non-uniform ECA can also be expressed by characteristic matrix, $T$. The matrix $T$ is a tri-diagonal matrix, where all elements except the elements of main, upper and lower diagonals are always zero (see Section~\ref{matrix}). 
In \cite{Das90c,Das91}, the matrix algebraic tool has been used for analyzing state transition behavior of this class of CAs.
From the matrix algebraic tool and characteristic polynomial, several interesting features of the non-uniform ECAs are derived. 

Apart from the seven linear/additive rules, there are another seven rules which are complement of the seven. These rules are named as complemented rules, see Table~\ref{CArule}. The CAs with these complemented rules can also be characterized by matrix algebra. In fact, a CA that uses the fourteen rules, seven linear and seven complemented, can be efficiently characterized by algebraic tools, see for example \cite{ppc1,NiloyFI08}\footnote{Some authors call the non-uniform CAs with linear/additive and complemented rules as additive CAs. The reason may be that, these CAs can be characterized using the tools of linear CAs. However, strictly speaking, these CAs are not additive, in general.}. However, for these CAs, we additionally need an {\em Inversion Vector} along with the characteristics matrix. An inversion vector $F$ for such an $n$-cell CA is defined as following:
\begin{equation}
  \begin{array}{l}
   \mbox{$F_i$ =}
    \left\{
	\begin{array}{ll}
	    \mbox{1,  if the $i^{th}$ cell uses a complemented rule}   ~~~~ \\
	    \mbox{0  otherwise}
	    \end{array}\right.
    \end{array}
 \end{equation}  
If a configuration $y$ of the CA is successor of an configuration $x$, then $y = T.x+F$.

Linear/additive CAs are broadly classified as group CAs and non-group CAs, in literature. Next, we briefly discuss about them.

\subsubsection{Group cellular automata}
\label{sscn_grpCA}
A non-uniform ECA is called a group CA if and only if the determinant $det (T) = 1$. The naming of the subclass of linear/additive non-uniform ECAs as group CAs comes from the fact that, these CAs form cyclic group ``under the transformation of operation with $T$'', see \cite{ppc1}. Group CAs are reversible CAs. 
In a group CA, therefore, all the configurations are reachable from some other configurations of the CA. 

In \cite{Pries86}, it is stated that if $\mathcal{R}_i$ is a {\em group rule} then its complement $\overline{ \mathcal{R}_i}$ (that is, $255 - \mathcal{R}_i$) is also a group rule, which is proved in \cite{Das90b}.
A subclass of group CAs is maximal length CAs, in which all non-zero configurations lie in the same cycle. As shown by many authors, such as \cite{Horte89c,Barde90,Serra90c}, these CAs produce pseudo-random patterns having high randomness quality, and are utilized in electronic circuit testing.

In case of maximal length CAs, the characteristics polynomial is primitive. \cite{cattell1996synthesis} have given us a scheme of synthesizing a maximal length CA from a given primitive polynomial over $GF(2)$. Rules 90 and  150 are used to construct such non-uniform ECAs, and no single rule can produce a maximal length CA. In \cite{Jetta95}, the maximal length CAs are synthesized up to size 500, where one or two cells follow rule $150$ and the rest follow rule $90$. Boundary condition of these maximal length CAs is null. However, \cite{Nandi96} have shown that, for a periodic boundary CA, the characteristic polynomial is factorizable; therefore, there exists no maximal length CA under periodic boundary condition.

\subsubsection{Non-group cellular automata}
\label{sscn_ngrpCA}

These are irreversible linear/additive non-uniform ECAs. Here, the characteristics matrix $T$ is singular, whereas the $T$ of a group CA is non-singular.
Non-group CAs are explored in different areas, see for example \cite{Bhatt95,Chakr93,santanu00,Rappid}. Any non-group CA is characterized by the following terms - 
\begin{itemize}
\item \emph{attractors}: cyclic states form attractors. If an attractor contains only a single state, it is said to be in graveyard state or a point state attractor or fixed point. 

\item \emph{$\alpha$-basin} or \emph{$\alpha$-tree}: the set of state(s) rooted at any attractor state $\alpha$, is termed as $\alpha$-basin or $\alpha$-tree. 

\item \emph{depth} or \emph{height} of a CA represents the number of single-step evolution, required by the CA to reach to the nearest cyclic state from a non-reachable (that is, Garden-of-Eden) state.
\end{itemize}

Some important findings about number of predecessors of the all-zero configuration and depth of non-uniform ECAs are reported in \cite{ppc1}.
A fraction of all reachable/non-reachable configurations of a uniform CA have been identified by \cite{Martin84a}. However, in general for any additive CA (uniform/hybrid), the fraction of reachable/non-reachable configurations can be numerated from the knowledge of the number of predecessors of a reachable state. Another scheme has been developed  by \cite{Acri08b} to identify and count reachable and non-reachable configurations of a finite (uniform/non-uniform) ECA, which can be linear and as well as non-linear.

Some fascinating classes of non-group CAs are also explored, like - multiple attractor CAs (MACAs) by \cite{santanu00,maji2003theory}, depth-$1*$ CAs (D1 $*$ CAs) by \cite{Chowd92d} and single attractor CAs (SACAs) by \cite{Das91}. These CAs have been employed in a broad range of purposes like hashing, see \cite{adcom00}, classification, see \cite{NiloyIV,maji2003theory}, designing easy and fully testable FSM, see \cite{Chowd93a}, etc.

\subsection{Non-Linear CAs}
\label{scn_nlCA}
During the early age of non-uniform ECAs, the non-linear non-uniform ECAs have not been widely explored, due to absence of proper characterization tool. In the past, several attempts have been made to study the characteristics of uniform CAs (linear or nonlinear) qualitatively and quantitatively in terms of parameters, such as $\lambda$ parameter (\cite{langton90}), $Z$ parameter (\cite{WuenscheRePEc,WuenscheI}) etc.
De Bruijn graph is also considered as a characterization tool of CAs (linear/nonlinear) (\cite{suttner91,JPovelliard}). 

A characterization tool, named {\em reachability tree} (see Section~\ref{tree}), has been proposed to characterize (finite) non-uniform ECAs \cite{SukantaTH,Acri08b,entcs/DasS09}. 
It has been shown that, out of 256 ECAs rules, only 62 rules can design a rule vector of non-uniform reversible ECA under null boundary condition, whereas for periodic boundary condition, additional 8 rules, that is, total 70 rules can take part in a rule vector of reversible automaton. However, the position of these rules in a rule vector is not arbitrary. For the purpose of getting a rule vector of reversible CA, the 62 rules are classified into six classes, which are related to each other. Table~\ref{nextclass} shows the classes and their relations. To get a rule vector of a reversible non-uniform ECA under null boundary condition, one needs to choose a rule from Table~\ref{first}, then the subsequent rules from Table~\ref{nextclass}, and finally the last rule from Table~\ref{last}. It can also be observed that the group CAs under null boundary condition can be decided from the tables. 

In \cite{NazmaTh}, reachability tree has been used to distinguish cyclic/acyclic configurations from the configuration space of non-linear non-uniform ECA. In the same work, $1$-D non-linear non-uniform ECA with point state attractors are also explored using reachability tree. However, \cite{Soumya2010,Soumya2011} have developed rule vector graph (RVG) for studying invertibility of $3$-neighborhood CA with null boundary condition.

{\small
\begin{table}
\caption{First and Last Rule Tables}
\centering
\subfloat[First rule table]{
\label{first}
\begin{tabular}{|c|c|}\hline
Rules for & Class of \\
${\mathcal R}_0$ & ${\mathcal R}_1$ \\\hline
 3, 12 &  I \\
 5, 10 & II\\
 6, 9 & III\\\hline
\end{tabular}
}
\subfloat[Last rule table]
{
\label{last}
\hspace*{0.75in}
\begin{tabular}{|c|c|}\hline
Rule class & Rule set\\
for ${\mathcal R}_{n-1}$ & for ${\mathcal R}_{n-1}$ \\\hline
 I & 17, 20, 65, 68 \\
  II & 5, 20, 65, 80 \\
 III & 5, 17, 68, 80 \\
 IV & 20, 65 \\
 V & 17, 68 \\
 VI & 5, 80 \\\hline
\end{tabular}
}
\end{table}
}
 
{\small
\begin{table*}
\caption{Class relationship of ${\mathcal R}_i$ and ${\mathcal R}_{i+1}$}
\begin{center}
\label{nextclass}
\begin{tabular}{|c|c|c|}\hline
Class of & ${\mathcal R}_i$ & Class of\\
${\mathcal R}_i$ & & ${\mathcal R}_{i+1}$ \\\hline
I & 51,  60,  195, 204 & I\\\cline{2-3}
  &                       85,  90,  165, 170 & II \\\cline{2-3}
  &     102, 105, 150, 153 & III \\\cline{2-3}
  &     53, 58, 83, 92, 163, 172, 197, 202 & IV \\\cline{2-3}
  &     54, 57, 99, 108, 147, 156, 198,201 & V \\\cline{2-3}
  &     86, 89, 101, 106, 149, 154, 166, 169 & VI \\\hline
II& 15,  30,  45,  60,  75,  90, 105, 120, 135, & I \\
   &                      150, 165, 180, 195, 210, 225, 240 & \\\hline
III& 15,  51, 204, 240 & I \\\cline{2-3}
   &                       85, 105, 150, 170 & II\\\cline{2-3}
   &    90, 102, 153, 165 & III \\\cline{2-3}
   &    23, 43, 77, 113, 142, 178, 212, 232 & IV\\\cline{2-3}
   &    27, 39, 78, 114,  141, 177, 216, 228 & V\\\cline{2-3}
   &    86, 89, 101, 106, 149, 154, 166, 169 & VI \\\hline
IV & 60, 195 & I \\\cline{2-3}
   & 90, 165 & IV \\\cline{2-3}
   & 105, 150 & V \\\hline
V  & 51, 204 & I\\\cline{2-3}
   & 85, 170 & II\\\cline{2-3}
   & 102, 153 & III\\\cline{2-3}
   & 86, 89, 90, 101, 105, 106, 149, 150, & VI\\
   &154, 165,166, 169 & \\\hline
VI & 15, 240 & I\\\cline{2-3}
   & 105, 150 & IV \\\cline{2-3}
   & 90, 165 & V \\\hline
\end{tabular}
\end{center}
\end{table*}
}

Non-linear non-uniform ECAs are proved to be efficient in various application fields, like VLSI design and test (\cite{SukantaTH}), pattern recognition and classification (\cite{MajiPhd,DasMNS09}), etc.
Further, designing of pseudo-random pattern generator (PRPG) around reversible non-linear non-uniform ECAs are reported in \cite{ats03,SukantaTH,tcad/DasS10}.

In \cite{NiloyV}, the concept of $GMACA$ (generalized MACA) has been introduced for non-linear non-uniform ECAs. The efficiencies of $MACA$ and $GMACA$ have been compared with respect to pattern recognition. In \cite{MajiPhd}, both linear and non-linear non-uniform ECAs have been used for designing a pattern classifier.

Fuzzy CA, a natural extension of boolean CA, is analyzed and synthesized using matrix algebraic tool by \cite{Maji05,DASFAA04}. This CA has also been used to design pattern classifier.

\section{Cellular Automata as Technology}
\label{scn_appCA}

Technology refers to the collection of techniques, methods or processes used to provide some services or solutions to problems, or in the accomplishment of an objective, such as scientific invigilation. The CAs have been historically used as a method for simulating biological and physical systems, and utilized to theoretically study such systems.
Since late $1980$s, however, the CAs have been started to be used as solutions to many real-life problems. In this section, we survey some of such solutions.

\subsection{Electronic circuit design}
The CAs, particularly non-uniform ECAs, have received their popularity as technology in the era of VLSI. The simplicity, modularity and cascadability of CA have enticed the researchers of VLSI domain. Some of these areas are briefly described here.

\subsubsection{Early phase developments:} CAs based machines, CAMs (CA Machines), having high degree of parallelism have been developed by \cite{Toffo87}, which are ideally suited for simulation of complex systems.
Even before the introduction of CAM, CAs have been utilized as parallel multipliers by \cite{Atrub65,Cole69}, parallel processing computers by \cite{Manni77}, prime number sieves by \cite{Fisch65}, and sorting machine by \cite{Nishi81}. Design of fault-tolerant computing machine by \cite{Nishio75} and  nanometer-scale classical computer by \cite{Benjamin97} are also commendable works. 

\subsubsection{VLSI Design and Test:}
\cite{Horte89a,Horte89c} has proposed non-uniform ECA based pseudo random pattern generator (PRPG) for built-in self-test (BIST) in VLSI circuits. Some of the major contributions in the research of PRPGs are reported in \cite{Das90b,Tsali91,Chowd92d,tcad/DasS10}. The CAs are also proposed as a framework for BIST structures by \cite{Tsali90,Das90b,Chowd92d,SukantaTH,DBLP:conf/ats/ChakrabortyC09} and as a deterministic test pattern generator by \cite{Albic87a,Das89,Das90b,SukantaTH}. Utilizing the scalability of non-uniform ECAs, a test solution for multi-core chips has been proposed in \cite{tcad/DasS10}.
While testing a CUT (Circuit-Under-Test) with pseudo-random patterns, a set of patterns may be prohibited to the CUT which may adversely affect the circuit. Non-uniform ECAs based solutions to this problem have been proposed by \cite{vlsi02a,ats03,SukantaTH} that can generate pseudo-random-patterns without PPS (Prohibited Pattern Set). Finally, a universal test pattern generator, termed as the $UBIST$ (Universal BIST) has been reported in \cite{Das93,ubist}. This $UBIST$ is able to generate any one of the four types of test patterns - (i) pseudo-random, (ii) pseudo-exhaustive, (iii) pseudo-random without PPS (Prohibited Pattern Set), and (iv) deterministic.

\subsubsection{Synthesis of Finite-State Machine (FSM):}
\cite{Mitra91b,Misra92b} have depicted the design of testable FSM with CAs. In \cite{Chowd93a,Chakr93}, some fascinating properties of a non-group CA and its dual, and their relationship are reported. These papers have also explored a particular class of non-group CAs, named as $D1*$CAs, which have been recommended ``as an ideal test machine which can be efficiently embedded in a finite state machine to enhance the testability of the synthesized design'', see \cite{Chakr93}.

\subsubsection{Security and others} A good number of works have been reported in literature that deal with CAs based encryption, cryptography and authentication techniques; see for example \cite{Nandi94a,Bao04,Seredynski2004753,ref1,DBLP:journals/jca/DasR11,DBLP:journals/ccds/DasC13,Formenti2014}. Other notable works, related to electronic circuit design are non-uniform ECAs based error correcting codes by \cite{Chowd94a,vlsi00b}, signature analysis by \cite{Horte90b,Das90e}, etc. For more discussion, see \cite{ppc1}.

\subsection{Computer vision and machine intelligence}
A group of researchers have also explored CAs in the fields of image processing, pattern recognition etc. These fields are roughly named as computer vision and machine intelligence.

\subsubsection{Image processing:} 

CAs, specially $2$-dimensional CAs, are used for image processing. They can address all the significant image processing works like translation, zooming, rotation, segmentation, thinning, compression, edge detection and noise reduction, etc., see for example \cite{Rosin06,Rosin2010790,Okba11}. 
\cite{khan98} has shown that using hybrid CAs, it is possible to rotate any image through an arbitrary angle. In \cite{Paul99}, a new GF($2$) CA based transform coding scheme for gray level and color still images has been proposed. Some works regarding edge detection and noise reduction are reported in \cite{Wongthanavasu03,Sadeghi12}.

\subsubsection{Pattern recognition:} 

CAs have been a popular tool for pattern recognition and classification since long, see for example \cite{Jen86,Raghavan93,ppc1}. As shown in the book by \cite{ppc1}, MACAs (multiple attractor CAs) can act as a natural classifier.
The correlation between MACA and \emph{Hamming Hash Family} ($HHF$) is shown in \cite{adcom00}. Hamming hash family has inherent capacity to address classification task. This basic framework of linear non-uniform ECA is extended to GMACA (generalized MACA) and utilized for modelling the associative memory, see \cite{Maji2,MajiPhd}. In \cite{DASFAA04}, fuzzy CAs have been explored as an efficient pattern classifier. Non-linear non-uniform ECAs based pattern classifiers have further been developed by \cite{DasMNS09}. Recently, an asynchronous CA based pattern classifier is reported by \cite{CPLX:CPLX21749}.

\subsubsection{Compression and others:} In \cite{Bhatt95}, some methods are proposed to perform text compression using CA as a technology. In \cite{Lafe}, cellular automata transforms are proposed for digital image compression and data encryption. It is shown that, CAs can generate orthogonal, semi-orthogonal, bi-orthogonal and non-orthogonal bases.
Non-uniform ECA based transforms have been presented by \cite{vlsi00a,ShawSM04,ShawDS06} for developing efficient schemes of image and document compression. Some other related works are reported in \cite{lafe2002method,ye2008novel}. A technology, called \emph{Encompression}, where encryption and compression are married, has been reported in \cite{Chandrama,ShawMSSRC04}.

\subsection{Medical science}

Since the seminal work of von Neumann in $1950$, CA has attracted attention of computer scientists and biologists as an excellent tool to model self-replicating system. The reason of choosing CA for biological modelling is -- ($1$) it is fast and easily implementable and ($2$) the visual result of the simulation provides remarkable resemblance with the original experiment. 
In \cite{Burks1984157}, the research on modelling the evolution and role of DNA sequences within the framework of CA has been initiated. \cite{ermentrout1993cellular} have reviewed a number of biologically motivated CAs, such as deterministic or Eurelian automata, lattice gas models and solidification models, heuristic CA etc. and shown the effectiveness of CAs in understanding a physical process.
In \cite{MitraDCN96}, a pioneering work on modelling amino acid using $GF(2^2)$ CA, called as amino acid CA (AACA), has been reported. Some other interesting works are recorded in \cite{zorzenon01,MOREIRA02,JCC:JCC20354,Santos:2013}.

CpG island detection in DNA sequence is a known problem in biological sequence analysis.
\cite{DBLP:conf/iicai/GhoshLMC07} has addressed this problem.  In \cite{DBLP:conf/acri/GhoshBMMC10}, a noteworthy class of non-uniform ECAs, known as \emph{Equal Length Cycle CAs}, has been proposed to predict and classify the enzymes. In \cite{Ghosh2012}, another notable class of CAs, termed as \emph{restricted $5$-neighborhood CA} (R5NCA) has been introduced to predict protein structure, and to develop a protein modelling CA machine (PCAM). Here, a R5NCA rule is used to model an amino acid of a protein chain. This PCAM has been used for synthesis of protein structure using an organized knowledge base, see \cite{DBLP:conf/acri/GhoshMC14}.

\section{Conclusion}
\label{scn_con}

Cellular automaton has gone through a long journey from the early period of von Neumann to elementary form of Wolfram, and to the modern trends of research using this simple yet beautiful model.
In this survey, various milestones of development regarding CAs are briefly depicted, such as the variety of types, different characterization tools, global behavior and non-uniformity.

During this journey of CAs, different types of automata have been developed by varying the parameters of CAs. Based on neighborhood dependencies of cells, various types of CAs have been developed. Similarly, varying dimension of cellular space, states per cell, local rules, we get various types of CAs.

In this survey we have discussed about characterization tools, such as de Bruijn graph, matrix algebra, reachability tree, etc, which have been used to understand the behavior of CAs. The global behavior of CAs, which are results of simple local interactions, are very interesting. We have surveyed some of the behavior, such as universality, reversibility, number conservation, computational ability, etc. Still, there are few more behavior of CAs, such as language recognition, which are not covered in this survey.

Then, we have focused our discussion on the non-classical CAs - asynchronous CAs, automata networks, and non-uniform CAs. We have noted that, in recent past, a good attention have been put on these non-classical CAs. Among them, the most explored non-classical CAs are (finite) non-uniform ECAs, where the local rules are always ECAs rules. We have dedicated a section for these CAs.

Lastly, we have discussed about the potential of CAs as technology. CAs have been used as solutions to many real-life problems. In fact, in this era of VLSI, the CAs, which can efficiently be implemented in hardware, can take a lead role to solve many real-life problems.

This survey is a brief history of cellular automata. We have tried to arrange a tour to different corners of the developments. As a result, many concepts are stated informally, and the writing style remains primarily narrative. However, a good number of references have been added in support of the presented concepts. Interested readers may go to these references to explore their interests.


\begin{acknowledgements}
The authors gratefully acknowledge the anonymous reviewers for their comments and suggestions, which have helped to improve the quality and readability of the paper.
\end{acknowledgements}

\bibliographystyle{spbasic}      
\bibliography{Ref}   


\end{document}